\documentclass[a4paper,aps,pre,showpacs,twocolumn]{revtex4}
\usepackage[dvips]{graphicx}
\usepackage{amsmath}
\usepackage{amssymb}
\usepackage{psfrag}
\usepackage{color}
\usepackage{natbib}
\usepackage{soul}
\usepackage{wasysym}
\usepackage{pifont}

\newcommand{\beq}{\begin{equation}}
\newcommand{\eeq}{\end{equation}}

\newcommand{\epsstrain}{\varepsilon}
\newcommand{\glv}{\gamma_{lv}}

\newcommand{\Tinf}{\glv}

\newcommand{\q}{\theta}

\newcommand{\upd}{\mathrm{d}}

\newcommand{\Ai}{\mbox{Ai}}

\newcommand{\lc}{\ell_c}
\newcommand{\lcurv}{\ell_{\mathrm{curv}}}

\newcommand{\Li}{L_I}
\newcommand{\Lo}{L_O}
\newcommand{\tLi}{\tilde{L}_I}
\newcommand{\tLo}{\tilde{L}_O}

\newcommand{\Lwrink}{L_w}
\newcommand{\Rfilm}{R_{\mathrm{film}}}
\newcommand{\Rnd}{{\cal R}}
\newcommand{\rtip}{r_{\mathrm{tip}}}

\newcommand{\mNT}{m_{\mathrm{NT}}}
\newcommand{\LNT}{L_{\mathrm{NT}}}

\newcommand{\rhol}{\rho_{\ell}}

\newcommand{\ksub}{K_{\mathrm{sub}}}
\newcommand{\keff}{K_{\mathrm{eff}}}
\newcommand{\ktens}{K_{\mathrm{tens}}}
\newcommand{\kcurv}{K_{\mathrm{curv}}}

\newcommand{\Ustrain}{U_{\mathrm{strain}}}
\newcommand{\Ustrainp}{U_{\mathrm{strain}}^{(\mathrm{planar})}}
\newcommand{\Ustrainb}{U_{\mathrm{strain}}^{(\mathrm{bare})}}
\newcommand{\ustrain}{u_{\mathrm{strain}}}
\newcommand{\Ubend}{U_{\mathrm{bend}}}

\newcommand{\Ugrav}{U_{\mathrm{gpe}}}
\newcommand{\ugrav}{u_{\mathrm{gpe}}}
\newcommand{\Usurf}{U_{\mathrm{surf}}}

\newcommand{\rmu}{{\rm u}}
\newcommand{\wo}{\delta}
\newcommand{\woo}{\delta_{\mathrm{mod}}}

\newcommand{\twoc}{\tilde{\delta}_c}
\newcommand{\tdelta}{\tilde{\delta}}
\newcommand{\tdeltaast}{\tilde{\delta}_\ast}
\newcommand{\tdeltaastast}{\tilde{\delta}_{\ast\ast}}
\newcommand{\tdeltaTrans}{\tilde{\delta}_{\mathrm{break}}}
\newcommand{\sqq}{\sigma_{\theta\theta}}
\newcommand{\srr}{\sigma_{rr}}

\newcommand{\tzeta}{\tilde{\zeta}}
\newcommand{\tpsi}{\tilde{\psi}}
\newcommand{\tF}{{\cal F}}
\newcommand{\tr}{\tilde{r}}
\newcommand{\tsqq}{\tilde{\sigma}_{\theta\theta}}
\newcommand{\tsrr}{\tilde{\sigma}_{rr}}
\newcommand{\awrink}{\tdelta_{\mathrm{compress}}}

\begin{document}
\title{Regimes of wrinkling in an indented floating elastic sheet}

\author{Dominic Vella$^1$ and Benny Davidovitch$^2$}
\affiliation{$^1$Mathematical Institute, University of Oxford, Woodstock Rd, Oxford OX2 6GG, United Kingdom\\
$^2$Physics Department, University of Massachusetts, Amherst, Massachusetts 01003, USA}

\begin{abstract}
A thin, elastic sheet floating on the surface of a liquid bath wrinkles when poked at its centre. We study the onset of wrinkling as well as the evolution of the pattern as indentation progresses far beyond the wrinkling threshold. We use tension field theory to describe the macroscopic properties of the deformed film and show that the system passes through a host of different regimes, even while the deflections and strains remain small. We show that the effect of the finite size of the sheet ultimately plays a key role in determining the location of the wrinkle pattern, and obtain scaling relations that characterize the number of wrinkles at threshold and its variation  as the indentation progresses. Some of our predictions are confirmed by recent experiments of Ripp \emph{et al.} [arxiv: 1804.02421].
\end{abstract}



\maketitle

\section{Introduction}

It is a natural human instinct to test an object's stiffness by poking it with one's finger.  Locally indenting the skin covering fruit, flesh or a (musical) drum gives a ready assessment of  the tension of the skin and of the pliability of the material. In precisely the same way, controlled indentation is used in a range of scientific applications to measure the internal pressure within viruses \cite{HernandoPerez12}, bacteria \cite{arnoldi00}, growing plant cells \cite{Milani13} and yeast cells \cite{arfsten10,Vella12} and to measure the stiffness of the polymeric capsules used in drug delivery \cite{gordon04}. In many of these examples, poking is intended to give a measure of the background tension in the object prior to indentation. However, the very act of poking can alter that state of tension and is even the basis of the technique often used to measure the stretching modulus of ultra-thin materials such as Graphene \cite{Lee2008} and Molybdenum Disulfide \cite{CastellanosGomez2015}. Furthermore,  depending on the  boundary conditions, indentation is able to turn tension to compression in some regions of the body.  If the poked body is a thin elastic sheet then such a compression can lead to  wrinkling \cite{Holmes10,Vella11,Vella15}, because the object's resistance to compression is extremely low.

\begin{figure}
\centering
\includegraphics[width=0.95\columnwidth]{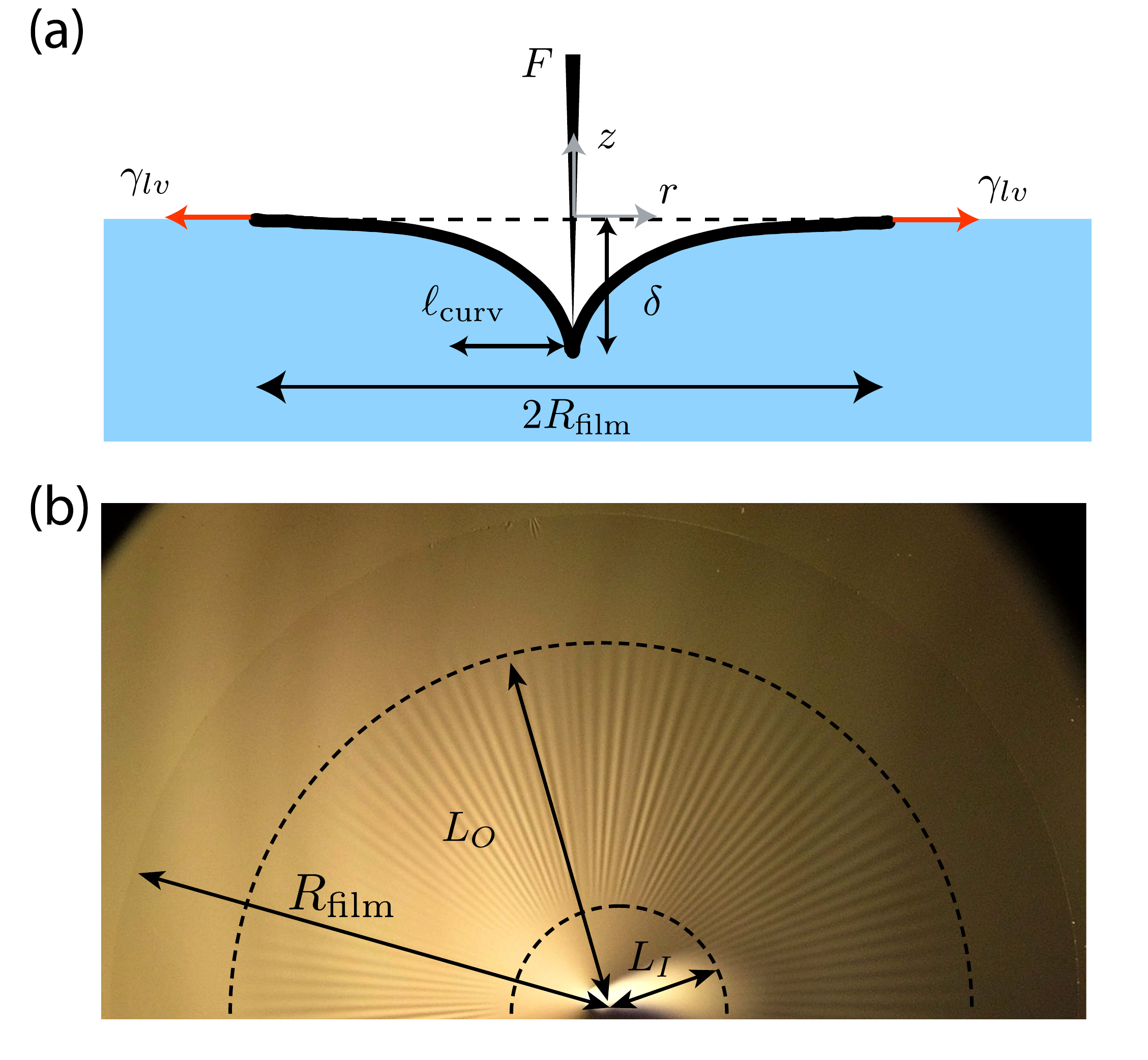}
\caption{The indentation of a floating elastic sheet by a vertical depth $\delta$ requires the application of a force $F$. (a) Schematic showing the experimental setup and key geometric quantities. (b) Experimental image (courtesy of Jiangshui Huang) showing the plan view of a poked, floating Polystyrene film ($\Rfilm=11.4\mathrm{~mm}$, $t=105\mathrm{~nm}$). Note that for this indentation depth, wrinkles occupy an annular region $\Li\leq r\leq\Lo<\Rfilm$, so that the film is in the intermediate wrinkling regime, denoted FT-II here.
}
\label{fig:Setup}
\end{figure}

One system that has been studied experimentally in some detail in recent years is the indentation of an elastic sheet floating at the surface of a water bath \cite{Holmes10,Vella15,Paulsen16,Box17,Ripp18},  shown schematically in fig.~\ref{fig:Setup}a. The tension of the air--water interface causes this sheet to be in a uniform and isotropic state of tension prior to indentation. The sheet is then indented by a small probe (radius $100\mathrm{~\mu m}\lesssim \rtip\lesssim500\mathrm{~\mu m}$). Above a critical indentation depth, the sheet is observed to develop radial wrinkles, which propagate (upon increasing indentation depth, see fig.~\ref{fig:Setup}b) and eventually  cover the whole sheet; ultimately the sheet transitions to a ``folded" or ``crumpled" state. These pattern-forming properties have been studied in some detail, with particular focus on the wrinkle number (or wavelength) that has led to new understanding of the pattern selection mechanisms in such systems \cite{Cerda03,Huang07,Davidovitch11,Davidovitch12,Paulsen16,Taffetani17}. Each of these previous studies has tended to focus on a different regime of the indentation: Box \emph{et al.}~\cite{Box17} studied small indentation depths of relatively thick sheets while several groups have studied larger indentation depths for ultra-thin (\emph{i.e.}~nanometre-thick) sheets \cite{Holmes10,Vella15,Paulsen16,Ripp18}. Here, we present a unified perspective on the different stages of such experiments, discussing how for a solid, Hookean sheet, the interplay between gravity, surface tension and elasticity, together with geometric nonlinearity, gives rise to a surprisingly rich and complex behavior.

We explore the deformation of the sheet, incorporating wrinkling via the Far-from-Threshold approach \cite{Davidovitch11}, and show that the gross shape is governed by three emergent length scales. Two of these  length scales, which we denote $\Li$ and $\Lo$, are the inner and outer horizontal extents of the annular wrinkled zone, respectively (see fig.~\ref{fig:Setup}b). The third emergent length scale is the characteristic radius of the portion of the sheet that is vertically deflected by poking, which we denote $\lcurv$ (fig.~\ref{fig:Setup}a). We show that these three length scales are coupled and, further, that they evolve as the indentation depth $\delta$ increases. This evolution and coupling of emergent length scales causes the system to pass through several qualitatively different regimes, each of which we elucidate through scaling arguments and detailed solutions of the F\"{o}ppl-von-K\'{a}rm\'{a}n (FvK) equations. Crucially, we find that, after wrinkles are formed, the deflected portion of the sheet grows upon further increase of the indentation depth, $\lcurv\sim \delta^{1/2}$. The growth of this deformed zone means that the finite size of the sheet will, ultimately, play a role in the deformation and consequently in the mechanical response (\emph{i.e.}~in determining the force $F(\delta)$ required to impose an indentation depth $\delta$). We then summarize previous results \cite{Vella15,Paulsen16} that addressed the final regime in which the outer boundary of the wrinkles coincides with the edge of the film.

\section{Physical Picture of Different Regimes \label{sec:physics}}

As the imposed indentation depth increases, it is unsurprising that the state of stress and deformation profile within the sheet should also change. More surprising is the delicate manner in which the stress and shape change during a single indentation experiment --- the transformation of the balance between surface tension, elasticity and gravity caused by changes in geometry leads  the sheet through a number of qualitatively different types of response. To give a sense of the physics underlying these different types of response, we present in this section an informal, physically based, view of the different regimes of indentation. Much of this discussion will focus on the different scaling laws that are observed in each regime and their origin. In \S\ref{sec:detcalcs} we shall present the equations that describe the sheet's morphology and state of stress (the F\"{o}ppl-von-K\'{a}rm\'{a}n, or FvK, equations) along with numerical solutions of these equations. This analysis confirms the scaling laws given in this section (up to some logarithmic factors) and, together with the asymptotic results of \S\ref{sec:asy},  also furnishes the appropriate numerical pre-factors to complement the scaling laws obtained in this section.

\subsection{The early stages of indentation}

Consider an elastic sheet of radius $\Rfilm$ and thickness $t$ floating at the surface of a liquid with surface tension coefficient $\glv$ and density $\rhol $ (the weight of the sheet itself is assumed negligible). The sheet has Young modulus $E$ and Poisson ratio $\nu$ so that its stretching stiffness $Y=Et$ and bending stiffness $B=Et^3/[12(1-\nu^2)]$. Prior to indentation, the stress within the sheet is isotropic, uniform and equal to the surface tension of the exterior liquid, \emph{i.e.}~$\sqq=\srr=\glv$, just as it would be for a fluid membrane with no shear rigidity. For small indentation depths, the elastic stresses caused by deformation are negligible in comparison to this uniform and isotropic pre-tension so that $\srr,\sqq\sim\glv$ in all but a small region close to the point of indentation. The equation of vertical force balance for the vertical displacement $\zeta(r)$ of the sheet may be written
\beq
 B\nabla^4\zeta- (\sigma_{rr} \kappa_{rr}  + \sigma_{\theta\theta} \kappa_{\theta\theta}) =-\ksub \zeta - \frac{F}{2\pi r} \delta(r), 
 \label{eqn:vfbalance}
\eeq where $F$ is the indentation force, which we assume for simplicity to be point-like, and $\ksub=\rhol  g$ is a foundation stiffness representing the restoring force due to the hydrostatic pressure within the liquid. The principal curvatures $\kappa_{rr}=\partial^2\zeta/\partial r^2$ and $\kappa_{\theta\theta}=r^{-1}\partial\zeta/\partial r+r^{-2}\partial^2\zeta/\partial \theta^2$.

With a uniform stress state $\srr,\sqq\sim\glv$ and negligible bending stiffness $B=0$, \eqref{eqn:vfbalance} is simply the Laplace--Young equation for the axisymmetric meniscus around a cylinder \cite{lo1983}. Sufficiently far from the point of indentation the solution of this equation  takes the form $\zeta\sim K_0(r/\lc)$ where $\lc=(\glv/\ksub)^{1/2}$ is the \emph{capillary length} and $K_0(x)$ is a modified Bessel function of the second kind \cite{abramowitz64}. Crucially, this solution demonstrates that, in this limit, the curvature of the surface  decays over a horizontal length scale comparable to the capillary length, \emph{i.e.}~$\lcurv\sim\lc$, just as would be the case for poking a bare liquid--gas interface  \footnote{Close to the point of indentation, it is known that the solution of the Laplace--Young equation depends sensitively on the curvature of the indenting object \cite{lo1983}. We do not consider this in detail since, in any case, the stress in an elastic sheet changes from the isotropic $\srr,\sqq\sim\glv$ as the indentation point is approached.}. Furthermore, we note that in this regime the work done by the indenter, $F\wo$, goes into the gravitational potential energy of the liquid that is displaced vertically (which at a scaling level is that of a cone of height $\wo$ and radius $\lcurv$). We therefore have $F\wo\sim \wo^2\rhol  g\lcurv^2$ and hence $F\sim\rhol  g\lcurv^2 \wo\sim\glv\wo$. Note that the work done, $F\delta$, is an even power of $\delta$: pushing down on the surface requires the same work as pulling up on it by the same amount.

The deformation of a fluid membrane is accurately described by Eq.~\eqref{eqn:vfbalance} with $\srr=\sqq=\glv$. However, for an elastic sheet the radial and hoop stresses $\srr(r)$ and $\sqq(r)$ are not known \emph{a priori}. Determining these stresses as the indentation increases involves a nonlinear coupling between the normal force balance equation, \eqref{eqn:vfbalance}, and the $2^{nd}$ FvK equation, the compatibility of strains. This coupling is described in detail in \S \ref{sec:detcalcs}, but a simpler view relies on geometry and shows that the elastic strain induced by indentation is
\beq
\epsstrain\sim (\delta/\lcurv)^2.
\label{eqn:geomstrain}
\eeq (Note that the indentation-induced strain \eqref{eqn:geomstrain} is quadratic in the indentation depth $\wo$: pushing down on the sheet induces the same strain as pulling up on it by the same amount!)  This strain induces a stress, $Y \epsstrain$, which, crucially, becomes comparable to the underlying stress in the film, $\glv$, when $\wo\sim \woo=\lc(\glv/Y)^{1/2}$ (since, for small indentations, $\lcurv\sim\lc$). For $\wo\gtrsim\woo$ the stress state within the elastic film is modified from its initial, isotropic state. To determine these modifications precisely requires a detailed numerical analysis; we therefore introduce a  dimensionless parameter to describe the indentation depth relative to $\woo$:
\beq
\tdelta=\frac{\wo}{\woo}=\frac{\wo}{\lc}\left(\frac{Y}{\glv}\right)^{1/2}.
\eeq

The dimensionless indentation depth $\tdelta$ will be a key parameter in this problem. Using $\glv$ and $\lc$ as the natural scales for stresses and lengths, respectively, we shall show in \S\ref{sec:detcalcs} that the problem also has two intrinsic dimensionless parameters, independent of the degree of indentation
\beq
\epsilon=B/(\glv\lc^2),\quad \Rnd=\Rfilm/\lc.
\label{eqn:NDpars}
\eeq

In \eqref{eqn:NDpars} $\epsilon$ is the reciprocal of the bendability introduced by \cite{Davidovitch11} (and should be distinguished from the typical strain, $\epsstrain$, of \eqref{eqn:geomstrain}). For the experiments on PS films described by Ripp \emph{et al.}~\cite{Ripp18}, $\epsilon\lesssim 10^{-4}$ so that the effects of the bending stiffness are negligible  in determining the radial profile of the film.  However, it is also known that, despite its small value, $\epsilon$ has a dramatic effect on the wrinkling pattern that forms with increasing indentation \cite{Paulsen16}, due to the efficient mechanism provided for an (almost) total relaxation of compressive stress.

The parameter $\Rnd$ in \eqref{eqn:NDpars} represents the dimensionless film radius. In the experiments reported by Ripp \emph{et al.}, $4\lesssim\Rnd\lesssim15$; we shall see that, provided $\Rnd\gg1$, its role may be neglected in the early stages of indentation but that, at sufficiently large indentation depths, the finite size of $\Rnd$ does play an important role.

Together with the parameters $\epsilon$ and $\Rnd$, the parameter $\tdelta$ is crucial for understanding the poking of an elastic sheet. We emphasize that none of these  parameters is encountered in the corresponding problem for a fluid membrane. From a mathematical perspective, it is the combination of all three that  makes the poking of an elastic sheet a much richer phenomenon than the poking of a fluid membrane.

The limit of small indentations, or linear response,  that we have just studied corresponds to the limit $\tdelta\ll1$. Ultimately we wish to study large indentations, \emph{i.e.}~$\tdelta \gg 1$ but subject to the assumption that the slope of the deflected shape remains small ($|\nabla \zeta| \ll 1$) so that the vertical force balance equation \eqref{eqn:vfbalance} remains valid. However, first we study the intermediate case where the indentation depth is neither large nor small, that is $\tdelta=O(1)$.

\subsection{$\tdelta=O(1)$ and the onset of wrinkling\label{sec:OnsetWrinkles}}

For intermediate values of the dimensionless indentation depth $\tdelta$, the stress profile within the film is not constant on the horizontal length scale $\lcurv$. As might be expected, the radial stress $\srr$ decreases monotonically to $\glv$ at the edge. However, the hoop stress $\sqq$ does not decrease monotonically with $r$. Rather, $\sqq$ decreases more quickly than $\srr$ and overshoots the far-field value $\sqq=\glv$ before increasing again \cite{Vella15}. Intuitively, this overshoot occurs because the tensile radial stress pulls material circles inwards towards the point of indentation; since the circumference of such circles is larger than that associated with their new position, these material circles become relatively compressed. Eventually, at a critical value $\tdelta=\awrink$ the hoop stress becomes compressive, $\sqq(r)<0$, within an annular region $\Li<r<\Lo$. Generally, one would expect the axisymmetric, unbuckled state of the poked sheet to become unstable for some $\tdelta = \tdelta_c(\epsilon)>\awrink$. However, the very thin films of interest here have exceedingly  small bending moduli ($\epsilon\ll1$), and so can only support a small level of compression before buckling: in the limit of very small bending resistance, $\epsilon \to 0$, this critical indentation depth to induce buckling $\tdelta_c(\epsilon) \to \awrink$. 

Once $\tdelta$ exceeds the critical value, a ``ring" of small width emerges, in which the hoop compression leads to a buckling instability and, hence, the  emergence of wrinkles in a narrow annulus; the inner radius of this annulus is $\sim\lc$ (as seen in fig.~\ref{fig:Setup}b) and its width, the length of the wrinkles, $\Lwrink=\Lo-\Li$.  In the wrinkled annulus, the shape is well described by: 
\begin{equation}
\zeta(r,\theta) \approx \zeta_0(r) + f(r)\cos(m\theta)  \ ,  
\label{eq:wrinkle-pattern}
\end{equation}
where the wrinkle amplitude $f(r) \ll \delta$. The wrinkle number, $m$, which may also vary spatially, defines a local wrinkle wavelength, $\lambda = 2\pi r/m$. 

Experiments \cite{Vella15,Ripp18} confirm that wrinkling occurs at a threshold indentation depth $\wo_c$ and, further, that the wrinkles are indeed confined to a narrow annular region. Based on the above scaling arguments, we expect that for highly bendable films ($\epsilon\ll1$)
\beq
\delta_c\sim \lc (\glv/Y)^{1/2}
\label{eqn:OnsetScaling}
\eeq with the pre-factor in this scaling being simply $\awrink$. Numerical solutions of the full problem, presented in \S\ref{sec:detcalcs}, suggest that the pre-factor $\awrink$ depends on the film size $\Rnd$ (see fig.~\ref{fig:RegimeDiag}) but that for $\Rnd\gg1$, $\awrink\approx11.75$. Previous experiments \cite{Vella15} confirm the scaling in \eqref{eqn:OnsetScaling} as the sheet thickness and liquid surface tension are varied. Further, the observed pre-factor is close to the value $\awrink\approx11.75$ predicted previously \cite{Vella15}, indicating that the explicit effect of bending stiffness on the wrinkling threshold is limited.

\subsubsection{Scaling behavior at wrinkle onset}

\begin{figure}
\centering
\includegraphics[width=0.9\columnwidth]{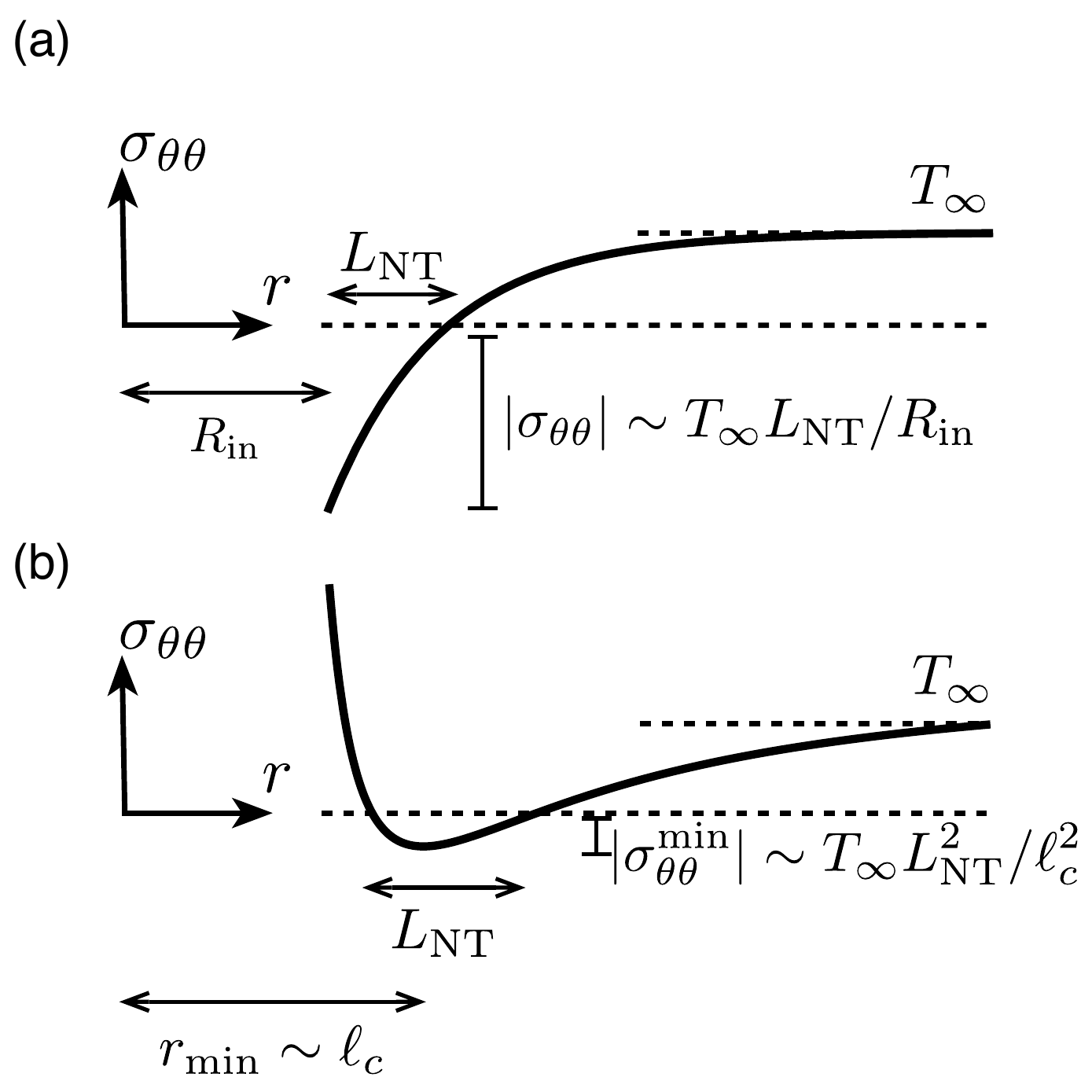}
\caption{The difference in the scaling of wrinkle wavenumber in the Near Threshold regime between  the wrinkling of an elastic annulus ($R_{\mathrm{in}}<r<\infty$) subject to distinct tensile loads at its boundary (the Lam\'{e} problem \cite{Davidovitch11}) and indentation-induced wrinkling in a complete sheet ($0<r<\infty$) may be understood by comparing the stress profiles in each case.  (a) In the Lam\'{e} problem the  hoop stress $\sqq$ is approximately linear as it becomes compressive, so that the depth of the minimum stress $|\sqq^{\mathrm{min}}|\sim L_{\mathrm{NT}}$. (b) For indentation,  $\sqq(r)$ is parabolic close to its minimum and so $|\sqq^{\mathrm{min}}|\sim L_{\mathrm{NT}}^2$. (In each case a far-field tension $T_\infty$ is applied; $T_\infty=\glv$ here.)
}
\label{fig:NTscalingSketch}
\end{figure}

To determine the scaling behavior of the wrinkle number and onset indentation depth with $\epsilon$, we follow the scaling approach of ref.~\cite{Davidovitch11} exploiting the expectation that the number of wrinkles $m\gg1$ for $\epsilon\ll1$. We shall also assume that the wrinkle length $\Lwrink=\LNT\ll\lc$ (with the subscript `NT' denoting that we are `Near Threshold'). These assumptions allow us to estimate that the dominant part of the bending term in \eqref{eqn:vfbalance}, $B\nabla^4\zeta\sim Bm^4\zeta/\lc^4$ (since the wrinkle amplitude vanishes outside an annulus of width $\LNT$ positioned at $r\sim \lc$), while the radial stretching term $\srr\partial^2\zeta/\partial r^2\sim \Tinf\zeta/\LNT^2$. The balance between these two terms gives $\LNT\sim (\Tinf\lc^4/Bm^4)^{1/2}$. However, the wrinkle number $m$ is still undetermined at this stage.

To make further progress, we must also estimate the size of the dominant portion of the azimuthal compression in \eqref{eqn:vfbalance}, which is $\sqq\kappa_{\theta\theta}\sim\sqq r^{-2}\partial^2\zeta/\partial \theta^2\sim \sqq m^2\zeta/\lc^2$. To this end, we note that $\sqq(r)$ necessarily has a parabolic profile close to its minimum, as shown in fig.~\ref{fig:NTscalingSketch}b. This suggests that the depth of the minimum, $|\sqq^{\mathrm{max}}|\sim \Tinf(\LNT/\lc)^2$ (assuming that the relevant horizontal length scale, namely the radius of the narrow annulus at which wrinkles emerge,   is again $\lc$). We therefore find that $m^6\sim\Tinf^2\lc^4/B^2$ or
\beq
\mNT\sim  \epsilon^{-1/3}.
\label{eqn:mNT}
\eeq We emphasize that the $\epsilon^{-1/3}$ scaling of \eqref{eqn:mNT} contrasts with the `Near Threshold' scaling $\mNT\sim \epsilon^{-3/8}$ that characterizes the Lam\'{e} problem \cite{Davidovitch11,Coman07}. This difference in scaling is a result of the parabolic approximation used above, rather than the simple linear approximation for the stress profile that is appropriate for that geometry \cite{Davidovitch11} (compare the hoop stress profile of fig.~\ref{fig:NTscalingSketch}a with that in fig.~\ref{fig:NTscalingSketch}b).

The final piece of the puzzle is to determine the `over-indentation' required to induce wrinkling (\emph{i.e.}~the amount of indentation beyond the value at which the hoop stress first becomes compressive). The amount of over-compression is expected to increase linearly with $\tdelta-\awrink$, \emph{i.e.}~$|\sqq|\sim\glv(\tdelta-\awrink)$ and so,  using the parabolic approximation of the stress profile (fig.~\ref{fig:NTscalingSketch}), we find that $\LNT^2\sim(\twoc-\awrink)\lc^2$. Combining this with the above estimates of $\LNT$ and the wrinkle number $\mNT$ we find that
\beq
\twoc(\epsilon)-\awrink\sim \epsilon^{1/3}.
\label{eqn:deltacNT}
\eeq

\subsubsection{Beyond onset}

In the preceding analysis, we have assumed that $\tdelta$ is only slightly beyond the threshold of wrinkling; in this case, the compressed annulus is so narrow and the compression is so weak, that the state of the sheet can be described by a small perturbation to the compressed, axisymmetric (unbuckled) state of the poked sheet. We call this parameter regime ``Near Threshold" (NT).  Although the NT behavior is the natural arena for the application of standard buckling theory \cite{Coman07}, it is crucial to understand that its experimental relevance \cite{Vella15,Paulsen16,Ripp18} is marginal, since in the high bendability regime ($\epsilon\ll1$) the NT regime occupies a small portion in the parameter space spanned by $\tdelta,\Rnd$ and $\epsilon$. In particular, the size of the NT region of parameter space vanishes as $\epsilon \to 0$. Beyond the NT regime, the wrinkled state of the poked sheet cannot be described as a perturbation to the compressed, unbuckled state. Instead, wrinkling relaxes the azimuthal compression to leave a compression-free stress field with radial tension only and no shear or hoop components --- this is the Far-from-Threshold (FT) regime \cite{Davidovitch11,Davidovitch12,Taffetani17} and is characterized by a hoop compression $\sqq(r)<0$ that vanishes as $\epsilon \to 0$. The relaxation of hoop stress leads to a ``slaving" relation between the wrinkle amplitude $f(r)$, wrinkle number $m$, and the inward radial displacement, $\rmu_r(r)$, which reads \cite{Davidovitch11} : 
\begin{equation}
\frac{f^2m^2}{4r^2} = - \frac{\rmu_r}{r} \ . 
\label{eq:slaving}
\end{equation} 
The above relation (written for simplicity for the case of Poisson ratio $\nu=0)$, expresses the fact that the length ``wasted" by azimuthal undulations matches the excess length of a hoop of radius $r$ that is displaced to $r + \rmu_r(r)$.   

\subsection{The Far-from-Threshold regime\label{sec:scalingFT}}

With radial tension as the only stress component, in-plane force balance, $\nabla\cdot\boldsymbol{\sigma}=0$, yields that $\srr(r) = C/r$ in the wrinkled region, $\Li<r<\Lo$. Continuity of radial stresses at the edges of the wrinkled zone, which is necessary for force balance, then gives:          
\beq
C = T_I\Li = T_O\Lo,
\label{eq:FT-A}
\eeq where $T_I =\srr(r=\Li)$ and $T_O =\srr(r=\Lo)$ are the radial stresses at the inner and outer edges of the wrinkled annulus, respectively.

Substituting $\srr(r) = C/r, \sqq=0$ into Eq.~\eqref{eqn:vfbalance}, the vertical force balance on the sheet reduces to Airy's equation \cite{Vella15,Paulsen16} so that the film shape $\zeta(r)$ in the wrinkled annulus $\Li<r <\Lo$ is an Airy function \cite{abramowitz64},
\beq
\zeta(r) \sim \Ai (r/\lcurv),
\label{eqn:MainTextAiry}
\eeq where  
\beq
\lcurv = (C/\ksub)^{1/3}
\label{eq:FT-B}
\eeq is the horizontal length scale over which the interfacial deformation decays; we shall see that, in this case,  $\lcurv\neq\lc$.
Finally, we note that the outer, unwrinkled annulus $\Lo<r<\Rfilm$, and the inner disk $r<\Li$ are under pure tension. This implies that the radial stress $T_O  = \sigma_{rr}(\Lo)$ is  a finite fraction of the surface tension $\glv$ that pulls on the boundary $r=\Rfilm$ \cite{Davidovitch11}. We therefore have that 
\beq 
T_O \sim \glv  \ . 
\label{eq:FT-C}
\eeq Meanwhile, the geometry that gave rise to the estimate of the indentation-induced strain \eqref{eqn:geomstrain} still holds; the typical geometrically induced stress is thus $T_{\mathrm{geom}}\sim Y(\delta/\lcurv)^2$. Assuming that the deflected sheet is approximately triangular, similar triangles then shows that the radial stress remains approximately constant in the deformed part of the film and, hence, that the inner stress at the inner edge of the wrinkle pattern $T_I=T(\Li)\sim T_{\mathrm{geom}}$. We may therefore write
\beq
T_I\sim Y(\wo/\lcurv)^2.
\label{eq:FT-D}
\eeq

Equations~\eqref{eq:FT-A}-\eqref{eq:FT-D} provide four independent scaling relations that couple the three emergent lengthscales $\Li,\Lo,\lcurv$, with the two emergent stresses $T_I,T_O$. We therefore need one more equation to complete our scaling analysis. This last equation distinguishes three sub-regimes of the FT behavior, which we denote by FT-I, FT-II, and FT-III in Table I. 
\vspace{0.5cm}

{\underline{\emph{FT-I}:}} Let us consider first the parameter regime $\tdelta \sim O(1)$. Here the wrinkled annulus is still concentrated close to the radius $\lcurv \sim \lc$; hence both the inner and outer radii of the wrinkle pattern are close to $\lcurv$ and so the wrinkled annulus is very narrow ({\emph{i.e.} $(\Lo-\Li)/\Li \ll 1$). As a result, various scaling laws are satisfied in a trivial manner: $\Li \sim\Lo\sim\lcurv\sim\lc$ and $T_I\sim T_O\sim\glv$.  Nevertheless, the pattern already exhibits features that reflect its strong deviation from the perturbative, NT regime. The most striking  signature of this difference is the number of wrinkles $m$; we have seen at a scaling level that  $m_{\mathrm{NT}}\sim \epsilon^{-1/3}$; however, experimentally a  new, geometry-dominated scaling behavior emerges that holds throughout the entire FT regime \cite{Paulsen16}.

\vspace{0.5cm}

{\underline{\emph{FT-II}:}} Let us address now the parameter regime $\tdelta \gg 1$, where the tension in the film is expected to be much larger than $\glv$ and so the characteristic radii of the wrinkled annulus should be much larger than the radius at onset, \emph{i.e.}~$\Li,\Lo\gg\lc$. Assuming first that the inner tensile zone ($r<\Li$) occupies a finite fraction of the vertically-deflected region ($r<\lcurv$), we can supplement Eqs~\eqref{eq:FT-A}--\eqref{eq:FT-D} with a final equation: 
\beq
\lcurv \sim  \Li.
\label{eq:FT-D1}
\eeq With this additional scaling relation, it is possible to determine the scaling relations for the  poking-induced tension $T_I$, and the radial distances $\Li,\Lo,\lcurv$ in terms of the depth of indentation, $\tdelta$. We find that: 
\begin{subequations}
\begin{gather}
\frac{\Li}{\lc} \sim \frac{\lcurv}{\lc} \sim \tdelta^{1/2} \\
\frac{\Lo}{\lc} \sim \tdelta^{3/2} \\ 
\frac{T_I}{\glv} \sim \tdelta 
\end{gather} 
\label{eq:FT-II-Scale}
\end{subequations}   

The scaling relations \eqref{eq:FT-II-Scale} exhibit a macroscopic feature of the floating film system that is fundamentally different to  the behavior of a fluid membrane: the lateral scale over which the vertical deformation of the film decays is no longer the capillary length $\lc$ but rather an emergent length scale $\lcurv\sim\lc\tdelta^{1/2}$, which increases as indentation progresses. Similarly, the tension at the inner boundary, $T_I$, depends on the indentation depth. Finally, the indentation force $F\sim\rhol  g\lcurv^2\wo\sim \glv\lc(\glv/Y)^{1/2}\tdelta^2$. We emphasize that this nonlinear response does not stem from a non-Hookean material response or from deflections with large slope (since our model assumes both Hookean elasticity and small slopes).  Rather, it is the geometrical nonlinearity encapsulated in the vertical force balance \eqref{eqn:vfbalance} combined with the indentation-induced stretching, which led to \eqref{eq:FT-D}, that is at its root.

The non-trivial geometric nonlinearity in the problem is also what leads to the surprising symmetry in the indentation force: one might expect the indentation force to be an odd power of the indentation depth since poking up by a distance $\delta$ will require a force of the same magnitude (but opposite sign) to poking downwards by the same distance. (Alternatively, the energy should be expected to be an even power of $\delta$ because of the up-down symmetry, as we saw for $\tdelta\ll1$.) However, the emergent length scale $\lcurv\sim\lc\tdelta^{1/2}$ complicates this picture:  this length scale is the same regardless of the sign of the indentation. More properly, one might write $\lcurv\sim\lc|\tdelta|^{1/2}$ so that $F\sim\glv\lc(\glv/Y)^{1/2}\tdelta|\tdelta|$. For simplicity, we shall not carry this modulus around in the following, and assume that if one wishes to study pushing from below ($\tdelta<0$), rather than pushing from above ($\tdelta>0$) then the correct sign is easily clarified by common sense.

Beyond the dependence of the indentation force on the poking depth, we note the nontrivial, explicit dependence of the characteristic radii and stresses in \eqref{eq:FT-II-Scale} on almost all of the control parameters: the surface tension, gravity, and stretching modulus of the sheet all enter via the dimensionless indentation depth $\tdelta$. Similarly, the energy of the system in this case, ${\cal U}\sim F\wo$, also depends on all of the control parameters in the system.

Finally, we note from Eq.~(\ref{eq:FT-II-Scale}b) that the outer edge of the wrinkle pattern grows with increasing indentation, $\tdelta$. Since the sheet is finite, the actual wrinkle pattern must terminate at $\Rfilm$ or $\Lo$, whichever is the smaller. Hence the scaling behavior, Eq.~(\ref{eq:FT-II-Scale}), must be limited to the range \mbox{$1 \ll \tdelta \ll \Rnd^{2/3}$}, as noted in Table I. 

\vspace{0.5cm}
{\underline{\emph{FT-III}:}}
Beyond the FT-II regime, \emph{i.e.}~when $\tdelta \gg \Rnd^{2/3} \gg 1$, the wrinkles must terminate at the edge of the sheet, i.e.
\beq
\Lo = \Rfilm \  , 
\label{eq:FT-D2}
\eeq   Combining this relation with the scaling relations Eqs.~\eqref{eq:FT-A}--\eqref{eq:FT-D} we find that: 
\begin{subequations}
\begin{gather}
\frac{\lcurv}{\lc} \sim \Rnd^{1/3} \\
\frac{\Li}{\lc} \sim \tdelta^{-2}\Rnd^{5/3} \\ 
\frac{T_I}{\glv} \sim \tdelta^2 \Rnd^{-2/3},
\end{gather} 
\label{eq:FT-III-Scale}
\end{subequations}

\noindent with the indentation force $F\sim \glv\Rnd^{2/3}\wo$. Surprisingly, the energy of the system in this case, ${\cal U}\sim F\wo\sim \glv\Rnd^{2/3}\wo^2$, is independent of the elastic properties of the sheet (the sheet's Young modulus $E$ and thickness $t$). This unusual behavior signifies that the sheet is able to deform with only a negligible amount of stretching: it is able to deform from a flat shape to a shape with non-zero Gaussian curvature because wrinkles (asymptotically) cover the whole sheet, allowing excess length to be `wasted' very cheaply --- bending is very easy for such thin films. In this final limit the inner position of the wrinkles $\Li\to0$ as the indentation depth grows --- the wrinkles asymptotically approach the indentation point at $r=0$,  in stark contrast to both the NT and the FT-II behaviors in which $\Li$ grows without bound as $\tdelta$ increases. Note that the scaling (\ref{eq:FT-III-Scale}b) is in good agreement with experimental data presented previously \cite{Vella15}, while the scaling for the indentation force has recently been verified by Ripp \emph{et al.} \cite{Ripp18}.

\subsection{Wrinkle wavelength in the FT regime}

For completeness we also consider, briefly, the wrinkle wavelength in the FT regime. It is well known \cite{Cerda03} that the preferred wrinkle wavelength, $\lambda$, for an elastic beam of bending stiffness $B$ sitting above a substrate of linear stiffness $\ksub$ is
\beq
\lambda=2\pi(B/\ksub)^{1/4}.
\label{eqn:LambdaLaw}
\eeq However, for well-developed, FT, wrinkling the energy argument that leads to \eqref{eqn:LambdaLaw} can be repeated with $\ksub$ replaced by an effective stiffness  that arises from tension and/or curvature along the wrinkles \cite{Cerda03,Paulsen16}. This effective stiffness $\keff$ may be written as the sum of three stiffnesses:
\beq
\keff=\ksub+\kcurv+\ktens
\label{eqn:Keff}
\eeq where $\kcurv=Y\kappa^2$ is a stiffness due to the curvature of the substrate along the wrinkles and $\ktens\sim C/r^3$ is a stiffness due to the tension along the wrinkles.

Since the curvature and tensional stiffness may, in general, vary spatially, eqn \eqref{eqn:LambdaLaw} implies that the wrinkle  wavelength should vary spatially also. This spatial variation has been observed both in regime FT-III of the present problem \cite{Paulsen16} and in numerical simulations of the indentation of a pressurized shell \cite{Taffetani17}. In both cases, a quantitative account of the observed wrinkle wavelength was obtained by using \eqref{eqn:LambdaLaw} with $\ksub$ replaced by the local (but spatially varying) value of $\keff$.

With the scaling analysis of the last subsection having determined the behavior of the various length scales $\Li$, $\lcurv$ and $\Lo$ as functions of the indentation depth in various asymptotic regimes, it is natural to present the corresponding scaling prediction for the wavelength $\lambda$ and/or wrinkle number $m\sim r/\lambda$. A summary of the main results is given in Table \ref{table:Mresults}. Here we make two notes about these scaling results. First, the FT scaling $m\sim\epsilon^{-1/4}$ (up to a pre-factor that is a function of $r$) should be contrasted with the NT scaling $m\sim \epsilon^{-1/3}$. As a result, increasing $\tdelta$ (while keeping $\epsilon$ fixed) leads to a reduction in the number of wrinkles (see also ref.~\cite{Davidovitch11}).  Secondly, notwithstanding its spatial variation, the wrinkle number $m\sim (\wo/t)^{1/2}$ throughout the sheet (in regime FT-II) and similarly  in the curved portion of the sheet (in regime FT-III): in other words, the wrinkle number is determined solely by the indentation depth measured relative to the thickness of the sheet (up to spatial variations). Finally, let us note that the dependence of the wrinkle number becomes more complicated near the inner and outer boundaries of the wrinkled zone, $\tr=\Li$ and $\tr=\Rnd$ in FT-III and $\tr=\Lo$ in FT-II, since it then depends on the size of the film, the surface tension etc.~\cite{Paulsen16}.

\subsection{Summary of scaling results}

In this section we have outlined the range of scaling laws that characterize the different regimes of the indentation of a floating elastic sheet. Our main results for the various emergent lengths ($\Li,\Lo$ and $\lcurv$), the two tensions $T_I,T_O$ and the indentation force $F$ are summarized in Table \ref{table:results} for ease of reference. 

The transition between regimes FT-II and FT-III gives a few interesting insights on the nature of the poking problem. Firstly, the radial distances $\Li$ and $\lcurv$ both increase with $\tdelta$ in regime FT-II. This behavior is  as would be expected even in the absence of wrinkling, where $\lcurv\sim \tdelta^{1/2}$, as reported previously \cite{Box17}. In fact, this non-linear scaling is a consequence of the geometrical nonlinearity associated with radially stretching the sheet by poking, rather than wrinkling \emph{per se}. A quantitative difference between the axisymmetric and wrinkled  cases is observed in FT-II for the propagation of the outer edge of the compressive region: $\Lo\sim\tdelta$ without wrinkling, while correctly accounting for wrinkling leads to $\Lo\sim\tdelta^{3/2}$. 

More noticeable, qualitative, changes occur when wrinkles reach the edge of the sheet and the system transitions to FT-III. Firstly, the lengths $\Li$ and $\lcurv$ become ``decoupled" because the length scale $\lcurv$, which characterizes the size of the vertically-deflected zone, saturates at a $\tdelta$-independent value, (\ref{eq:FT-III-Scale}a). Secondly, the value of $\Li$, which marks the beginning of the wrinkled zone, starts to decrease sharply upon increasing the poking amplitude $\tdelta$.  (This latter behavior is a direct result of wrinkling: since wrinkling ensures that $\Li=T_O\Lo/T_I=\glv\Rfilm/T_I$ in this regime, the increase in  tension $T_I$ from the increased indentation depth must be accompanied by a decrease in $\Li$.) 

Finally, it is interesting to note that the saturated, $\tdelta$-independent value of the horizontal length scale of the deflected portion of the film, $\lcurv\sim\lc^{2/3}\Rfilm^{1/3}$, in regime FT-III, together with the value $\lcurv\sim\tdelta^{1/2}$ found in the FT-II regime, are both markedly different from the analogous length $\lcurv = \lc$ of a poked fluid membrane. In contrast with what was observed in the FT-II regime, however, $\lcurv$ in the FT-III regime has  an explicit and strong dependence on the size $\Rfilm$ of the sheet. 

\begin{table*}
\centering
\caption{Scaling results for the gross, macroscopically measurable properties, of a poked ultra-thin sheet (\emph{i.e.}~excluding the microscopic wrinkle number). Here asymptotic results are summarized in three asymptotic regimes.  }
\begin{tabular}{@{\vrule height 10.5pt depth4pt  width0pt}llcccccc}
      & Regime & $\lcurv/\lc$ & $\Li/\lc$  & $\Lo/\lc$ &  $F (Y/\glv)^{1/2}/(\glv\lc)$ & $u_r(\Rfilm)$
 \\ \hline\hline
& $\tdelta/\tdelta_c \ll 1  $  & $1 $  & --- & --- &  $4\pi \tdelta/\log(1/\epsilon)$  & $\glv \Rfilm/Y$   \\ 
& (Below Threshold) & \mbox{} & \mbox{} &  \mbox{} & \mbox{} 
\\
\hline
      &$1 \ll \tdelta \ll  \Rnd^{2/3}$  & ${\tdelta}^{1/2}  $&$ \lesssim \tfrac{\lcurv}{\lc} $ & ${\tdelta}^{3/2} $
&  $\tdelta^2$ &  $\glv \Rfilm/Y$
 \\ 
& (FT-II) & (up to log)   &  & (up to log)
&  (up to log) &    \\
\hline
 & $\tdelta \gg  \Rnd^{2/3} $  & $\Rnd^{1/3}$ & 
$ \tdelta^{-2} \Rnd^{5/3} $ & $\Rnd$  
&  $\Rnd^{2/3}\tdelta$ & $\wo^2/\lcurv$
\\ 
& (FT-III)
\\ \hline \hline
\end{tabular}
\label{table:results}
\end{table*}

Before moving on to discuss a more quantitative analysis of the problem, we pause briefly to discuss our expectations of how the above scaling results might be modified in related indentation problems, specifically the indentation of a clamped sheet or a pressurized shell. For a clamped sheet, the negative (inward) radial displacement that underlies the onset of wrinkling does not exist (unless the clamping is such that only  out-of-plane deflections are prevented, while radial sliding is allowed); there is therefore no possibility of wrinkling. In the indentation of a pressurized shell, many of the key ingredients are similar to those in the current problem: an externally applied tension exists prior to indentation (a consequence of the internal pressure, which requires an internal tension to balance it) and a geometrical stiffness due to the shell's radius $R$ plays the role of the hydrostatic pressure, in particular $\ksub\to \kcurv=Y/R^2$. These ingredients give rise to a horizontal length scale $\ell_p\sim (pR^3/Y)^{1/2}$, relevant for small indentation depths. For larger indentation depths, the length $\lcurv\sim (\delta R)^{1/2}$ is the analogue of the result seen here in regime FT-II, though in this case $\Lo\sim\lcurv\sim (\delta R)^{1/2}$ and $\Li\sim\delta^{-1/2}$; nevertheless $F\sim\delta$. Previous studies \cite{Vella11,Vella12,Vella15EPL} have only focussed on shallow shells, for which wrinkles are prohibited from reaching the edge of the domain. Therefore, to our knowledge, a parameter regime analogous to regime FT-III has not yet been identified.

\begin{table*}
\centering
\caption{Asymptotic regimes for the wrinkle number expected to be observed at different radial positions, $\tr$,  in a poked  floating, ultrathin sheet. Scaling results are shown for each of the three different stiffnesses that may determine the wrinkle wavelength $\lambda$; that which dominates at a particular radial position is highlighted in bold. }
\begin{tabular}{@{\vrule height 10.5pt depth4pt  width0pt}l|c|c|c}
\hline

      & $r\sim\Li$ & $r\sim\lcurv$ & $r\sim\Lo$ \\ \hline
      \hline
  \underline{\emph{NT}} & & &\\
$m\sim \tr\lc/\lambda$ & $\epsilon^{-1/3}$ & $\epsilon^{-1/3}$ & $\epsilon^{-1/3}$\\  
      \hline
 \underline{\emph{FT-II}} & & &\\
$\ktens/(\rhol  g)$ &1 &1 &$\tdelta^{-3}$\\
$\kcurv/(\rhol  g)$ & 1& 1&$0$\\
$\ksub/(\rhol  g)$ &\textbf{1} & \textbf{1}&\textbf{1}\\
$m\sim \tr\lc/\lambda$ &$\epsilon^{-1/4}\tdelta^{1/2}\sim (\delta/t)^{1/2}$ & $\epsilon^{-1/4}\tdelta^{1/2}\sim (\delta/t)^{1/2}$&$\epsilon^{-1/4}\tdelta^{3/2}$\\
\hline
 \underline{\emph{FT-III}} & & &\\
$\ktens/(\rhol  g)$ &$\boldsymbol{(\tdelta/\Rnd^{2/3})^6}$ &$1$ &$\Rnd^{-2/3}$\\
$\kcurv/(\rhol  g)$ & $(\tdelta/\Rnd^{2/3})^2$& $\boldsymbol{(\tdelta/\Rnd^{2/3})^2}$&$0$\\
$\ksub/(\rhol  g)$ &1 & 1&\textbf{1}\\
$m\sim \tr\lc/\lambda$ &$\epsilon^{-1/4}\tdelta^{-1/2}\Rnd^{2/3}$ & $\epsilon^{-1/4}\tdelta^{1/2}\sim (\delta/t)^{1/2}$&$\epsilon^{-1/4}\Rnd$\\
 \hline
  \hline

\end{tabular}
\label{table:Mresults}
\end{table*}

\section{Detailed Calculations\label{sec:detcalcs}}

\subsection{The FvK equations}

The axially symmetric vertical deflection of an elastic film is denoted $\zeta(r)$ and satisfies the vertical force balance equation \eqref{eqn:vfbalance}, which may be written
\beq
B\nabla^4\zeta-\srr\frac{\upd^2\zeta}{\upd r^2 }-\sqq\frac{1}{r}\frac{\upd\zeta}{\upd r}+\rhol  g\zeta=-\frac{F}{2\pi} \frac{\delta(r)}{r},
\label{eqn:FvK1}
\eeq where $B=Et^3/[12(1-\nu^2)]$ and the components of the axially symmetric stress tensor in the elastic film, $\srr(r)$ and $\sqq(r)$, must satisfy the in-plane equilibrium of the solid, which requires
\beq
\frac{\upd}{\upd r}\left(r\srr\right)=\sqq. 
\label{eqn:stressEqm}
\eeq The means of solving \eqref{eqn:stressEqm} are qualitatively different, depending on whether the elastic film is completely tensile (\emph{i.e.}~$\srr,\sqq>0$ everywhere) or whether there is a region in which the film wrinkles, $\sqq<0$. We therefore discuss the solution of \eqref{eqn:stressEqm} in each of these two cases in \S\ref{sec:AxisMembrane} and \S\ref{sec:TFT}, respectively.

\subsubsection{Unwrinkled sheet shapes\label{sec:AxisMembrane}}

In unwrinkled regions of the sheet it is possible to eliminate the radial displacement $u_r$ to obtain an equation relating stress to the deflection of the film. This equation reflects the compatibility of strains and may be combined with the equilibrium equation to close the problem. A common simplification is to introduce an Airy stress potential $\psi(r)$ defined in such a way that equilibrium \eqref{eqn:stressEqm} is automatically satisfied, e.g.~choose $\srr=\psi/r$, $\sqq=\upd\psi/\upd r$. We adopt this strategy so that the equation of compatibility of strains becomes
\beq
r\frac{\upd }{\upd r}\left[\frac{1}{r}\frac{\upd }{\upd r}\left(r\psi\right)\right]=-\frac{1}{2}Y\left(\frac{\upd \zeta}{\upd r}\right)^2.
\label{fvk2:dim}
\eeq This equation may be combined with the equation of vertical force balance \eqref{eqn:FvK1} to determine the shape of the sheet, and the stress state within the sheet. As we discussed in \S\ref{sec:physics}, for sufficiently large indentation depths (corresponding to $\tdelta\gtrsim O(1)$), the hoop stress becomes compressive, $\sqq<0$, in an annular region $\Li\leq r\leq \Lo$. The appearance of wrinkles qualitatively changes the behavior of the sheet in these regions, and so a different approach is required. We discuss this now.

\subsubsection{Tension field theory \label{sec:TFT}}

In highly wrinkled regions, the compressive hoop stress is relaxed, \emph{i.e.}~$\sqq=0$ or, more precisely, the ratio $\sqq(r)/\srr(r)\to0$ as the bendability $\epsilon^{-1}\to\infty$ \cite{Pipkin86,Steigmann90,Davidovitch11}. Equation \eqref{eqn:stressEqm} then immediately gives that $\srr=\glv\lc C/r$ for some dimensionless constant $C$. This stress state can be directly substituted into the vertical force balance equation \eqref{eqn:FvK1} to determine the shape of the sheet in the wrinkled annulus $\Li<r<\Lo$ up to the constant $C$. Outside the wrinkled annulus, \emph{i.e.}~for $0<r<\Li$ and $\Lo<r<\Rfilm$ the shape of the film and the state of stress are again found by solving  \eqref{eqn:FvK1} and \eqref{fvk2:dim}.

\subsection{Non-dimensionalization}

The natural horizontal length scale in this problem is the capillary length $\lc$, while the natural vertical length scale is $\lc(\glv/Y)^{1/2}$. Similarly we will non-dimensionalize stresses by $\glv$. We shall denote dimensionless variables by $\tilde{~}$ and find that \eqref{eqn:FvK1} becomes
\beq
\epsilon\nabla^4\tzeta-\tsrr\frac{\upd^2\tzeta}{\upd \tr^2 }-\tsqq\frac{1}{\tr}\frac{\upd\tzeta}{\upd \tr}+\tzeta=-\frac{\tF}{2\pi} \frac{\delta(\tr)}{\tr}
\label{eqn:FvK1ND}
\eeq where $\tsqq=0$ within the wrinkled region, $\epsilon$ is as defined in \eqref{eqn:NDpars} and
\beq
\tF=\frac{F}{\glv\lc}\left(\frac{Y}{\glv}\right)^{1/2}.
\eeq

The dimensionless bending stiffness $\epsilon\ll1$ and so  we shall neglect the first term in \eqref{eqn:FvK1ND} for all but the small deflection behavior. (This comes at the expense of neglecting boundary layers that are located close to the point of indentation \cite{Vella17}.) We note that the bendability $\epsilon^{-1}=\tau^{2}$ with $\tau$ a dimensionless tension introduced by \cite{Box17}.

The dimensionless version of the compatibility of strains, \eqref{fvk2:dim}, is
\beq
\tr\frac{\upd }{\upd \tr}\left[\frac{1}{\tr}\frac{\upd }{\upd \tr}\left(\tr\tpsi\right)\right]=-\frac{1}{2}\left(\frac{\upd \tzeta}{\upd \tr}\right)^2,
\label{eqn:FvK2ND}
\eeq which is to be applied in tensile regions (\emph{i.e.}~regions in which the $\tsrr,\tsqq>0$); in regions where $\tsqq$ would otherwise become compressive, we instead use tension field theory, so that
\beq
\tsrr=\frac{C}{\tr},\quad \tsqq=0.
\label{eqn:srrWrink}
\eeq Note that equations \eqref{eqn:FvK1ND}--\eqref{eqn:srrWrink} are invariant under the reflection $\tzeta\to-\tzeta$, $\tF\to-\tF$: this is the up--down symmetry discussed in \S\ref{sec:physics}.

\subsection{Boundary and matching conditions}

In the solution of the dimensionless problem \eqref{eqn:FvK1ND} with \eqref{eqn:FvK2ND} in tensile regions and \eqref{eqn:srrWrink} in the wrinkled region, a number of boundary conditions are required. However, when there is a wrinkled annulus in the sheet, there are also a number of matching conditions between wrinkled and unwrinkled regions that must be solved.

\subsubsection{Purely tensile sheets}

Below the wrinkling threshold, the shape of the sheet and the state of stress within it may be determined by solving the dimensionless FvK equations \eqref{eqn:FvK1ND} (with $\epsilon=0$) and \eqref{eqn:FvK2ND} subject to the boundary conditions of a given vertical displacement and the initial (unpoked) horizontal displacement at the point of indentation, i.e.
\beq
\tzeta(0)=-\tdelta,\quad  u_r(0)=\lim_{\tr\to0}\bigl[\tr\tpsi'-\nu\tpsi\bigr].
\label{bcs:originND}
\eeq 

The boundary conditions at the edge of the film, $\tr=\Rnd$, depend on the nature of the experiment. It is natural to impose that the radial tension matches the surface tension of the bath, \emph{i.e.}~$\glv=\srr(\Rfilm)=\psi(\Rfilm)/\Rfilm$. The appropriate boundary condition on the vertical displacement at the edge of the film requires one to consider the effect of a liquid meniscus which may affect the fine details of a wrinkle pattern in the vicinity of the edge \cite{Huang10}. However, for the purpose of the current paper, we may safely ignore this effect, and take:    
\begin{eqnarray}
\tzeta(\Rnd)&=&0\nonumber\\
\quad\tsrr(\Rnd)=\frac{\tpsi(\Rnd)}{\Rnd}&=& 1.
\label{bcs:inftyND}
\end{eqnarray}  Note that we now have two second order ODEs to solve, and hence require four boundary conditions, which are given by \eqref{bcs:originND} and \eqref{bcs:inftyND}. When addressing the small-indentation response (\emph{i.e.}~$\tdelta\ll1$), we would like to compare also the effect of the bending rigidity, and hence will compare results of membrane theory (which assumes $\epsilon = 0$ in eqn \eqref{eqn:FvK1ND} for the axisymmetric deformation), with an analysis that assumes  $0 < \epsilon \ll1$. In this case, the vertical force balance equation becomes fourth order and so two further boundary conditions are required; these are simply that the sheet avoids a cusp or discontinuity in slope so that $\zeta'=0$ at both $\tr=\tr_{\mathrm{ind}}$ and $\tr=\Rnd$. 

The numerical solutions of the system \eqref{eqn:FvK1ND} and \eqref{eqn:FvK2ND} subject to the boundary conditions \eqref{bcs:originND} and \eqref{bcs:inftyND} may be found using, for example, the boundary value problem solver \texttt{bvp4c} in MATLAB.

\subsubsection{Partially wrinkled sheets}

When the solution of the purely tensile problem exhibits hoop  compression, $\tsqq<0$, in some annulus, the problem changes: there is then an unknown region, $\tLi<\tr<\tLo$, in which the stress state is given by \eqref{eqn:srrWrink}. To solve this wrinkled problem, we again use the boundary conditions \eqref{bcs:originND} and \eqref{bcs:inftyND}, but must also match the wrinkled region $\tLi<\tr<\tLo$ to the tensile regions $0<\tr<\tLi$ and $\tLo<\tr<\Rnd$.


The problem now contains $13$ unknowns ($4$ for each of the two unwrinkled regions, each of which consist of two coupled second-order ODEs, and $2$ for the wrinkled region, which is described by a single second-order ODE, as well as the constant $C$ and the two lengths $\tLi$ and $\tLo$). The $4$ boundary conditions in \eqref{bcs:originND} and \eqref{bcs:inftyND} must therefore be supplemented by matching conditions between the wrinkled and unwrinkled regions
\begin{eqnarray}
\tzeta(\tLi^-)&=\tzeta(\tLi^+),\quad\tzeta(\tLo^-)=\tzeta(\tLo^+) \label{eqn:match}\\
\tzeta'(\tLi^-)&=\tzeta'(\tLi^+),\quad\tzeta'(\tLo^-)=\tzeta'(\tLo^+)\\
\tpsi(\tLi^-)&=C,\quad\tpsi(\tLo^+)=C,\\
 \tpsi'(\tLi^-)&=0,\quad\tpsi'(\tLo^+)=0.
\end{eqnarray} This takes us to $12$ conditions in total. The final condition arises from noting that the presence of unwrinkled regions in $\tr<\tLi$ and $\tr>\tLo$ together with the continuity of both stresses at each of these boundaries gives $u_r(\tLi)=u_r(\tLo)$ \cite{Vella15,Vella15EPL}. We then use Hooke's law within the wrinkled region,
\beq
\frac{\glv\lc}{Y}\frac{C}{r}=\frac{1}{Y}\left(\srr-\nu\sqq\right)=\epsilon_{rr}=\frac{\partial u_r}{\partial r}+\frac{1}{2}\left(\frac{\partial \zeta}{\partial r}\right)^2.
\eeq to show that
\beq
2C\log\frac{\Lo}{\Li}=\int_{\tLi}^{\tLo}\left(\frac{\partial\tzeta}{\partial \tr}\right)^2~\upd \tr,
\label{eqn:epsrr}
\eeq finally closing the system. The problem \eqref{eqn:FvK1ND}-\eqref{bcs:inftyND} with matching conditions \eqref{eqn:match}--\eqref{eqn:epsrr} can be solved numerically using the multipoint boundary value problem feature of \texttt{bvp4c} in MATLAB. 

\subsection{Numerical results and the effect of wrinkling}

The two quantities of most practical interest are the force-indentation relationship (the force law) and the evolution of the limits of the wrinkle pattern, $\tLi$ and $\tLo$, as the indentation depth increases. Here, we present numerical results for the variation of these quantities. We are particularly interested in testing the scaling predictions of \S\ref{sec:physics}, including determining the effect of wrinkling on these quantities (\emph{i.e.}~a comparison of how the behavior of these quantities changes between axisymmetric membrane theory and tension field theory). For this comparison we consider   the sheet radius $\Rnd$ to be large enough that it does not influence the results; unless otherwise stated, the numerical results presented here have $\Rnd=10^3$.

\subsubsection{Indentation force}

Figure \ref{fig:ForceLawNums} shows how the numerically determined indentation stiffness, $k=\tF/\tdelta$, varies as the inverse bendability, $\epsilon$, and indentation depth, $\tdelta$, vary. In fig.~\ref{fig:ForceLawNums}a we see that the effect of the bendability is only relevant for the small indentation depth behavior: all of the curves with different bendabilities ultimately `join' the nonlinear membrane theory result. We will make this notion of the transition between different regimes more concrete shortly. However, at  indentation depths beyond the onset of wrinkling, we see that there is a small deviation between the result of the axisymmetric nonlinear membrane theory (dash-dotted curve) and the prediction of tension field theory (solid curve).

\begin{figure}
\centering
\includegraphics[width=0.95\columnwidth]{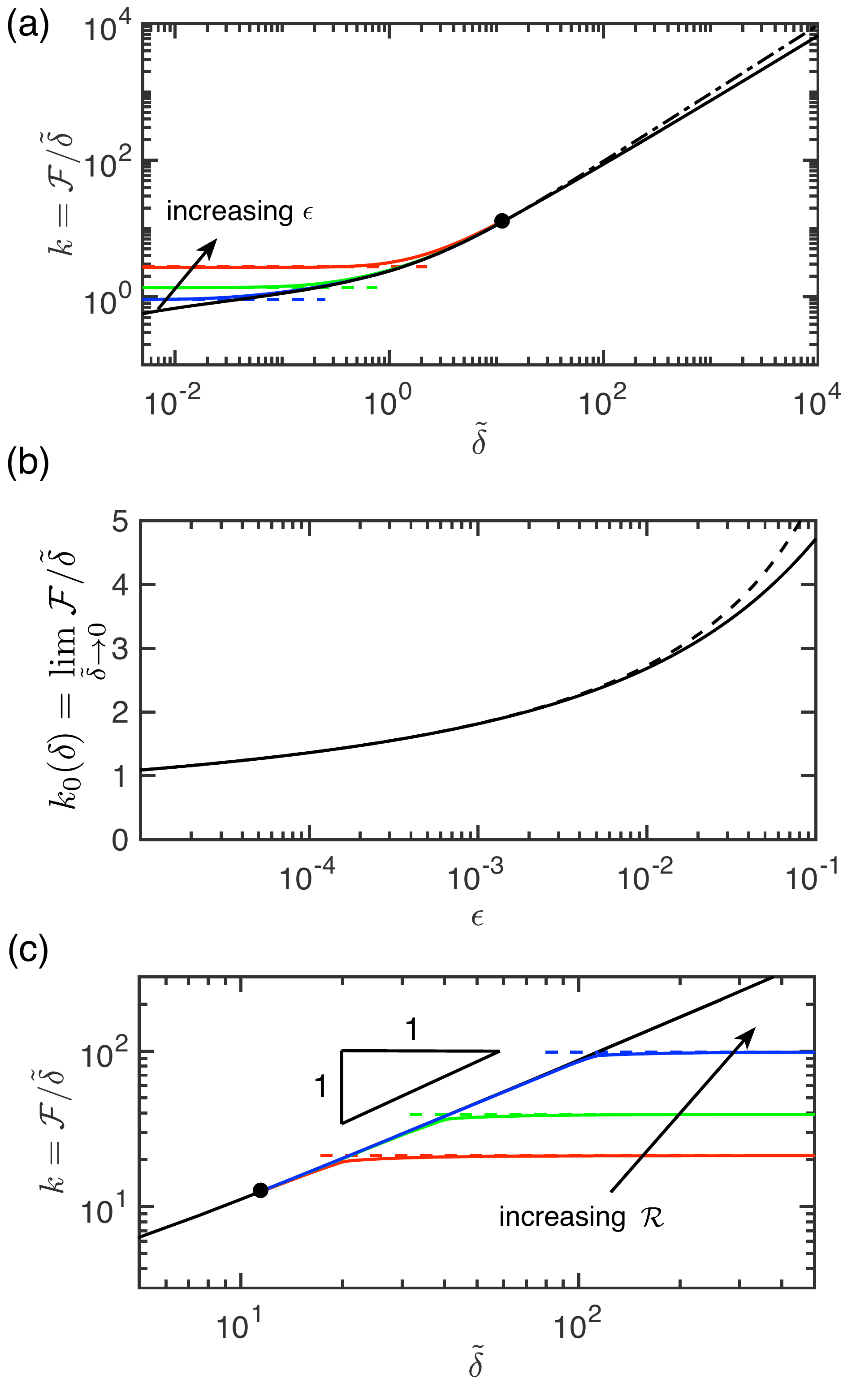}
\caption{(Color online) The indentation stiffness, $k(\tdelta;\epsilon)=\tF/\tdelta$. (a) $k(\tdelta)$  for a range of dimensionless bending stiffnesses $\epsilon$: $\epsilon=0$ (black), $\epsilon=10^{-6}$ (blue), $\epsilon=10^{-4}$ (green) and $\epsilon=10^{-2}$ (red); the direction of increasing $\epsilon$ is indicated by the arrow. The small indentation depth stiffness, $k_0=\lim_{\tdelta\to0}\tF/\tdelta$, for each finite value of $\epsilon$ is shown, calculated from \eqref{eqn:K1asy}, by the dashed horizontal line of the same color. The onset of wrinkling in membrane theory, $\tdelta=\awrink$, is shown by the solid black circle. For $\tdelta>\awrink$, axisymmetric membrane theory predicts that $\tF\sim\tdelta^2$ (dash-dotted curve) while tension field theory (accounting for the presence of wrinkles) predicts sub-quadratic growth.  (b) The variation of the small indentation linear stiffness $k_0(\epsilon)=\lim_{\tdelta\to0}k(\tdelta;\epsilon)$ with inverse bendability $\epsilon$. The full analytical expression \eqref{eqn:K1} (solid curve) is well described by the asymptotic result \eqref{eqn:K1asy} for $\epsilon\ll1$ (dashed curve), provided that $\epsilon\lesssim 10^{-2}$. (c) The indentation stiffness changes qualitatively as the wrinkles reach the edge of the sheet, transitioning from $k\sim\tdelta$ (up to logarithmic factors) to $k\sim\mathrm{~const}$. Numerical results with $\epsilon=0$ are shown as solid curves, with dashed lines give the corresponding asymptotic behavior \eqref{eqn:FT3Stiffness}, for $\Rnd=\Rfilm/\lc=10$ (red), $25$ (green) and $100$ (blue); the direction of increasing $\Rnd$ is indicated by the arrow.}
\label{fig:ForceLawNums}
\end{figure}

\subsubsection{Extent of the wrinkled region}

Our numerical analysis determines the extent of the region in which wrinkles occur: wrinkles are expected to occupy the annulus $\tLi<\tr<\tLo$. Both $\tLi$ and $\tLo$ are functions of the indentation depth, $\tdelta$, as shown in fig.~\ref{fig:WrinkleExtents}a. Also indicated in fig.~\ref{fig:WrinkleExtents}a are the regions in which the axisymmetric membrane theory predicts a compressive hoop stress, $\tsqq<0$. We note that in relaxing the compressive hoop stress, \emph{i.e.}~setting $\tsqq=0$ (as in tension field theory) rather than allowing $\tsqq$ to become arbitrarily negative (as in membrane theory), the extent of the wrinkled region grows considerably. Note that the numerically determined behavior of the inner position of the compressive region, $\tLi$, is consistent with the $\tdelta^{1/2}$ scaling predicted in \S\ref{sec:physics} in both cases: the presence of wrinkling  appears to only change the pre-factor. However, the outer limit $\tLo$ scales differently in the two cases: forcing the sheet to remain axisymmetric gives $\tLo\sim\tdelta$, while accounting for wrinkling appropriately appears to give $\tLo\sim\tdelta^{3/2}$. (We shall see later that there is, in fact a logarithmic term modifying these scalings.)

\begin{figure}
\centering
\includegraphics[width=0.95\columnwidth]{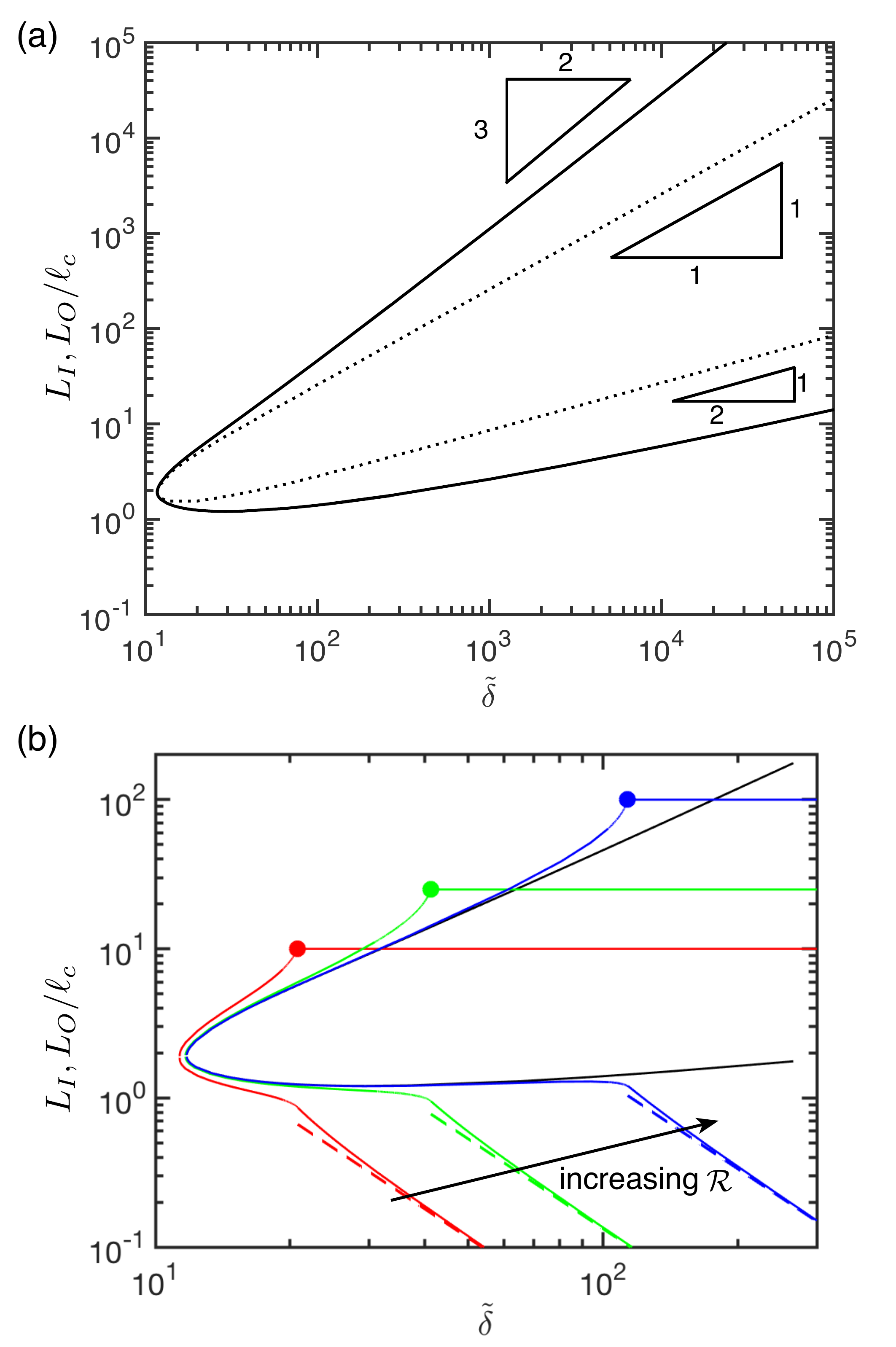}
\caption{(Color online) The wrinkled region is confined to an annulus $\tLi<\tr<\tLo$, which grows with indentation depth $\tdelta$.  (a) Numerical results for the wrinkle extents $\tLi$ and $\tLo$ as functions of the indentation depth $\tdelta$ for the case of an infinite sheet, $\Rnd=\infty$. The solid curve shows the results of the FT theory; the dotted curve shows the extent of the region of compressive hoop stress, $\tsqq<0$, according to axisymmetric membrane theory.  (b) For $\Rnd<\infty$, the evolution of $\tLi$ and $\tLo$ with indentation depth $\tdelta$ is significantly modified. Here results are shown for $\Rnd=10$ (red curves), $\Rnd=25$ (green curves) and $\Rnd=100$ (blue curves) with the results for an infinite sheet from (a) reproduced (solid black curve); the direction of increasing $\Rnd$ is indicated by the arrow. Note that the wrinkles reach the outer edge of the sheet, $\tLo=\Rnd$, at a finite indentation depth (denoted by a filled circle), which leads to the transition from FT-II to FT-III. This transition causes a significant change in the propagation of the inner boundary of the wrinkles, leading to the scaling $\tLi\sim\tdelta^{-2}$ as $\tdelta\to\infty$, shown as the dashed lines \cite{Vella15}. }
\label{fig:WrinkleExtents}
\end{figure}

\section{Asymptotic results \label{sec:asy}}

In this section we focus on understanding some of the features of the numerical results of \S\ref{sec:detcalcs} asymptotically. We present this discussion in a manner that mirrors the qualitative discussion of \S\ref{sec:physics}, \emph{i.e.}~we study how the stress, and consequently the shape and exerted force $F(\delta)$, change as the indentation depth $\tdelta$ changes.

\subsection{Small indentations}

In the absence of indentation, the stress within the sheet is homogeneous and isotropic: $\srr=\sqq=\glv$. For sufficiently small indentation depths, we do not expect the stress  to be significantly modified from this state, and so assume that $\tsrr\approx\tsqq\approx1$; the deflection $\tzeta(\tr)$ is then governed by the plate equation
\beq
\epsilon\nabla^4\tzeta-\nabla^2\tzeta+\tzeta=-\frac{\tF}{2\pi} \frac{\delta(\tr)}{\tr},
\eeq which may be solved analytically for axisymmetric deformations with a point indenter, $\zeta(r,\theta)=\zeta(r)$, as shown by \cite{Box17} (note that their dimensionless tension $\tau=\epsilon^{-1/2}$ in our notation). Box \emph{et al.} \cite{Box17} showed that in this regime the indentation force is linear  in the indentation depth, $\delta$, with
\beq
F= k_0(\epsilon) \times \glv \delta
\label{eqn:FversusD1}
\eeq where
\beq
k_0(\epsilon)=2\pi\frac{(1-4\epsilon)^{1/2}}{\mathrm{arctanh}\bigl[(1-4\epsilon)^{1/2}\bigr]}
\label{eqn:K1}
\eeq is a dimensionless spring stiffness.

Note that in \eqref{eqn:FversusD1} we have written the force--displacement response in a way that is particularly transparent for the limit of highly bendable sheets, $\epsilon\ll1$: the force is linearly proportional to both the surface tension coefficient and the indentation depth. However, we also note that even in this limit there remains a small dependence on the bending stiffness of the sheet since
\beq
k_0(\epsilon)\approx\frac{4\pi}{\log(1/\epsilon)}
\label{eqn:K1asy}
\eeq for $\epsilon\ll1$. For the experiments of Ripp \emph{et al.}, the inverse bendability $7\times10^{-9}\lesssim\epsilon\lesssim 7\times10^{-5}$; in this region of extremely bendable sheets, the asymptotic expression \eqref{eqn:K1asy} gives an extremely good account of the full analytical expression, \eqref{eqn:K1} (see fig.~\ref{fig:ForceLawNums}b).

\subsection{Transition from small to intermediate indentation}

\begin{figure}
\centering
\includegraphics[width=0.95\columnwidth]{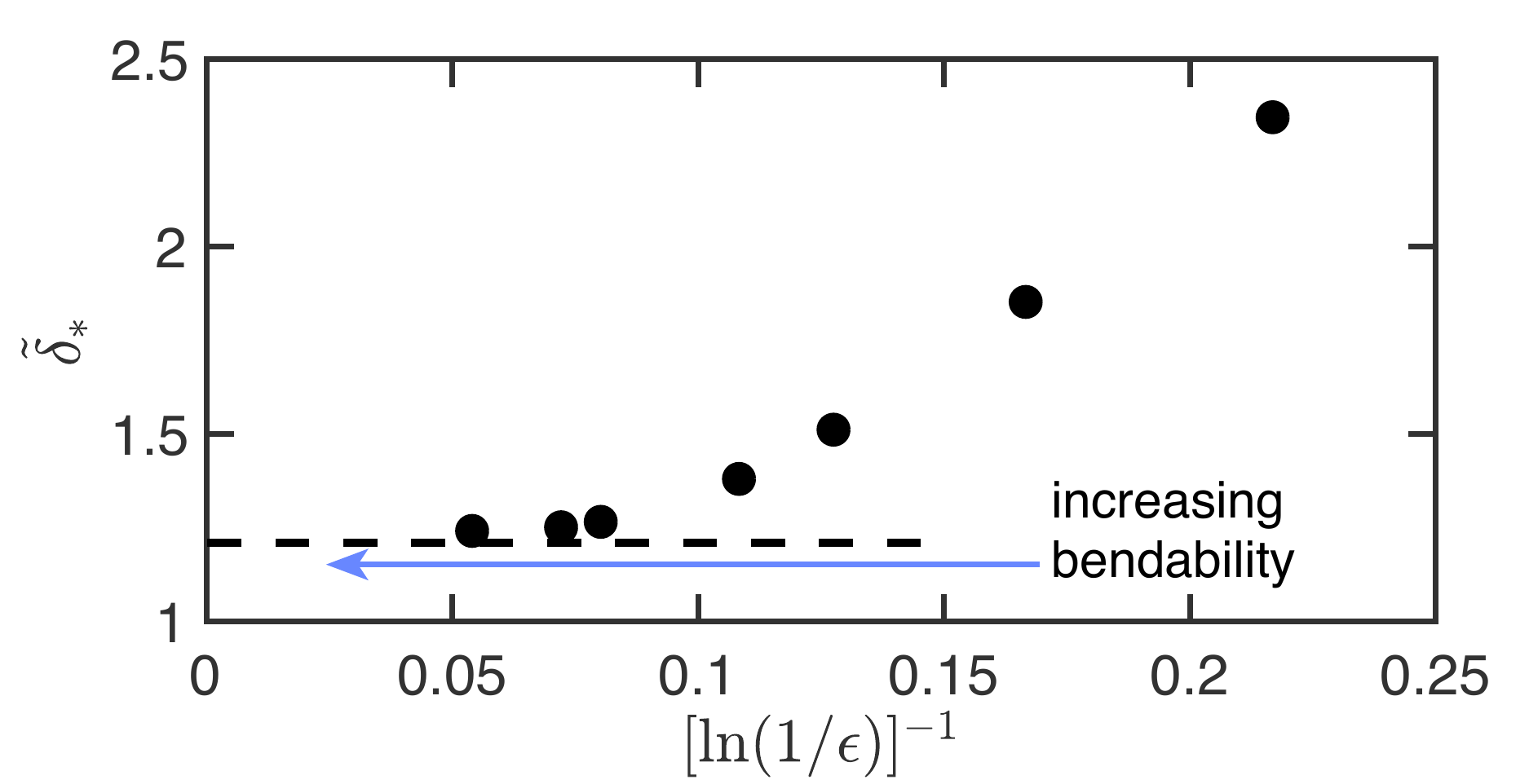}
\caption{The value of the indentation depth at the `transition' between constant and variable stiffness, $\tdeltaast$ defined in \eqref{eqn:DefnDeltaStar}, determined numerically for varying  inverse bendability, $\epsilon$ (points). Note that the value of $\tdeltaast$ varies by less than a factor of two over a wide range of bendabilities ($10^{-8}\lesssim \epsilon\lesssim 10^{-2}$). The value predicted by membrane theory $\tdeltaast(\epsilon=0)\approx1.21$ is shown by the dashed line. }
\label{fig:DeltaTransition}
\end{figure}

As the indentation depth $\tdelta$ grows, the indentation-induced stress grows in comparison to the `pre-tension' exerted by the liquid--vapor interface. In \S\ref{sec:scalingFT} we saw that this should lead to a change in the indentation force from a constant-stiffness mode (\emph{i.e.}~$\tF/\tdelta=k_0(\epsilon)$, with $k_0$ given in \eqref{eqn:K1}) to a stiffness that grows approximately linearly with  $\tdelta$ (\emph{i.e.}~$\tF/\tdelta\sim\tdelta$ for $\tdelta\gg1$). While we shall see shortly that the actual behavior for large indentation depths is slightly different than this expectation, the question of when this transition occurs is nevertheless of interest. Since the linear and nonlinear regimes are only asymptotic results, we should not expect there to be a precise transition point --- instead the transition between the two asymptotic results will occur gradually. However, a plot of $\tF/\tdelta^{3/2}$ will be non-monotonic with a turning point that is caused by this transition between linear and quadratic scalings. To make the notion of a  transition more concrete, and measurable, we therefore define
\beq
\tdeltaast:=\min_{\tdelta}(\tF/\tdelta^{3/2}).
\label{eqn:DefnDeltaStar}
\eeq We expect that this definition should be close to the definition introduced by Ripp \emph{et al.}~\cite{Ripp18}. We note that combining our definition with that used by \cite{Ripp18} we would expect $\tdeltaast\propto1/\log(1/\epsilon)$, based on \eqref{eqn:FversusD1}. The numerically determined value of $\tdeltaast$ is shown in fig.~\ref{fig:DeltaTransition} as a function of the inverse bendability.  We note that, while there seems to be a broad zone of high bendabilities ($\epsilon^{-1} \gg 1$) for which this expectation is consistent with the numerical results), for very large bendabilities the value of $\tdeltaast$ does not vary appreciably and, in particular, $\tdeltaast=O(1)$ even as $\epsilon\to0$.  This is because, even with $\epsilon=0$, a well-defined value of $\tdeltaast$ exists, corresponding to the transition from a sub-linear force law ($\tF\sim\tdelta/\log(1/\tdelta)$ for $\tdelta\ll1$ \cite{Vella17}) to a super-linear force law for $\tdelta\gg1$.

Note also that $\tdeltaast<\awrink$ (where wrinkles first appear) --- this cross-over from linear to quadratic force laws  occurs before the onset of wrinkling, as reported experimentally by Box \emph{et al.} \cite{Box17}. The onset of wrinkling is a more visible manifestation of the transition from small to large indentation depths and so we move on to discuss this transition now.

\subsection{The onset of wrinkling}

In the membrane theory presented previously \cite{Vella15}, it was found that the critical indentation depth at which the hoop stress becomes compressive, $\tsqq<0$, is $\twoc\approx11.75$ for an infinite sheet, $\Rnd=\infty$. In ref.~\cite{Vella15} it was expected that the highly bendable sheets considered, $\epsilon\ll1$, would wrinkle at values of the indentation depth very close to this  critical value (since they have very limited resistance to bending and so only have minimal resistance to compression). For real elastic films, however, there are two caveats that  limit the applicability of this result: (i) A finite radius may alter the critical value $\awrink$ of the indentation depth above which a compressive hoop stress appaers and (ii) A finite bendability allows the sheet to resist \emph{some} amount of compression before forming wrinkles. We would therefore expect that the critical indentation depth at which wrinkling is first observed is described by a function $\twoc=\twoc(\Rnd;\epsilon)$. In this section, we consider the effect of both finite bending stiffness and finite sheet radius. We first consider  the effect of finite sheet size to determining $\awrink(\Rnd)$. We then move on to understand the  scaling analysis of the critical indentation depth for finite $\epsilon$, but neglecting finite sheet size, \emph{i.e.}~we determine $\twoc(\Rnd=\infty,\epsilon\ll1)$.

\subsubsection{Finite sheet size, $\Rnd<\infty$}

The numerically determined threshold indentation depth for the onset of wrinkling in an infinitely bendable, but finite radius, elastic sheet is shown in fig.~\ref{fig:RegimeDiag}. This shows that for $\Rnd\gtrsim10$ the critical indentation depth at onset is within $10\%$ of the value for an infinite sheet reported by Vella \emph{et al.} \cite{Vella15}. However, for smaller values of $\Rnd$, the threshold decreases appreciably.

\begin{figure}
\centering
\includegraphics[width=0.95\columnwidth]{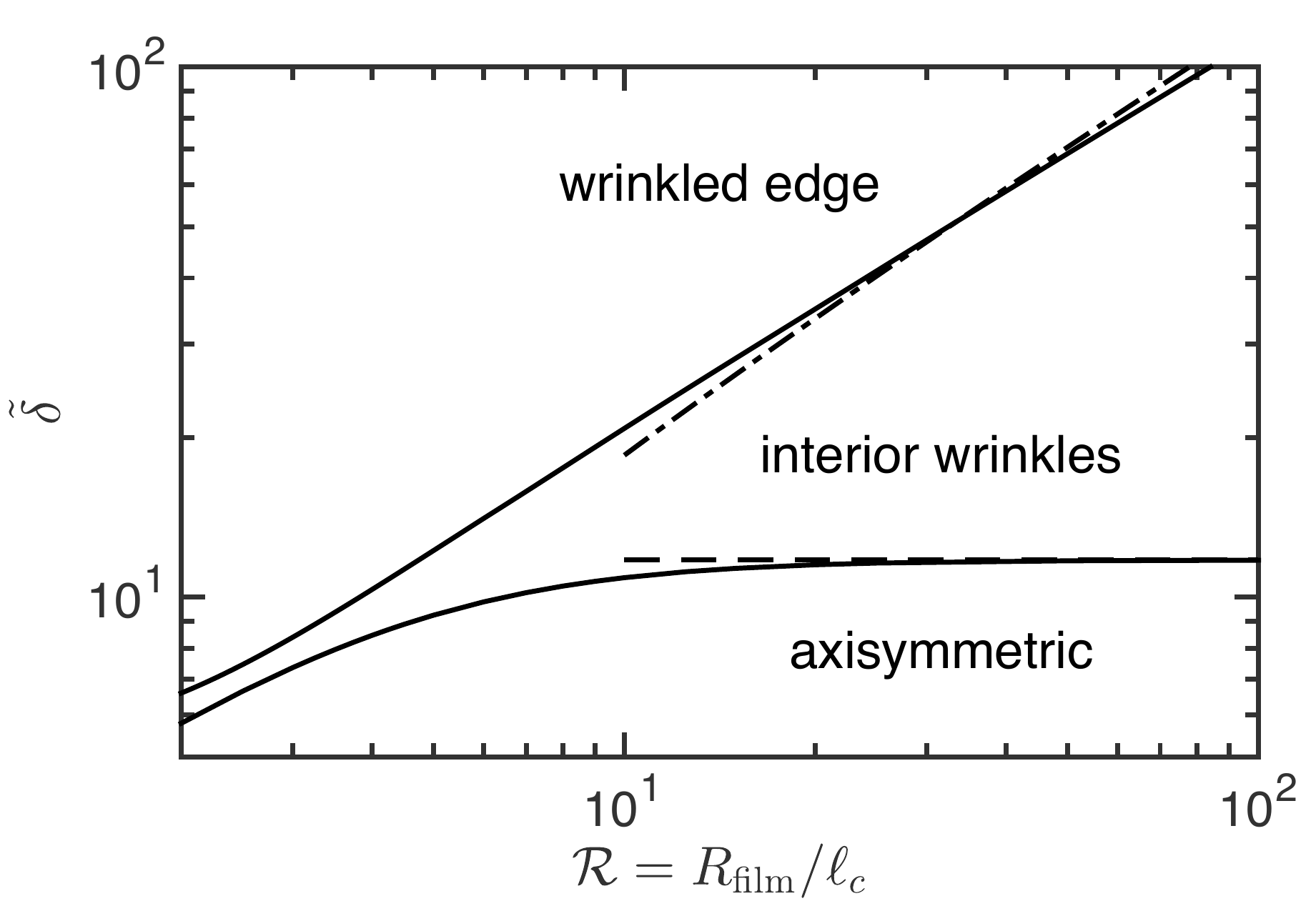}
\caption{Diagram showing how a film of finite radius but infinite bendability ($\epsilon=0$) behaves upon indentation. For a given film radius, the deformation is  axisymmetric at sufficiently small indentation depths, but wrinkles emerge at a critical indentation depth $\awrink(\Rnd)$ (the lower solid curve), which tends to the limiting value $\awrink(\infty)=11.75$, reported previously \cite{Vella15}  for infinitely large sheets $\Rnd=\infty$ (dashed horizontal line). At another critical indentation depth, $\tdeltaastast(\Rnd)$, the wrinkles reach the outer edge of the sheet. The upper solid curve shows the numerically determined  $\tdeltaastast(\Rnd)$, whilst the asymptotic result \eqref{eqn:dcritFTIItransAsy} is shown by the dash-dotted line. }
\label{fig:RegimeDiag}
\end{figure}

\subsubsection{Finite bendability, $\epsilon>0$}

To test the scalings for the properties of the wrinkle pattern at onset, \eqref{eqn:mNT}--\eqref{eqn:deltacNT},  we reanalyse previously published numerical results \cite{Box17}. These numerical results are (re-)plotted in fig.~\ref{fig:NTnums} (note that a different non-dimensionalization of the indentation depth $\wo$ was used in ref.~\cite{Box17}; their $\tau=\epsilon^{-1/2}$). Figure \ref{fig:NTnums}b shows that the `over-indentation', $\twoc-\awrink$, is linear in $\epsilon^{1/3}$ (at least for sufficiently small $\epsilon$). Similarly, fig.~\ref{fig:NTnums}c shows that the wrinkle number scales with $\epsilon^{-1/3}$, as expected. While these scalings have been observed in numerical results previously \cite{Taffetani17,Box17}, the argument given in \S\ref{sec:OnsetWrinkles} is, to our knowledge, the first exposition of the relevant balances that lead to  this scaling.

\begin{figure}
\centering
\includegraphics[width=0.95\columnwidth]{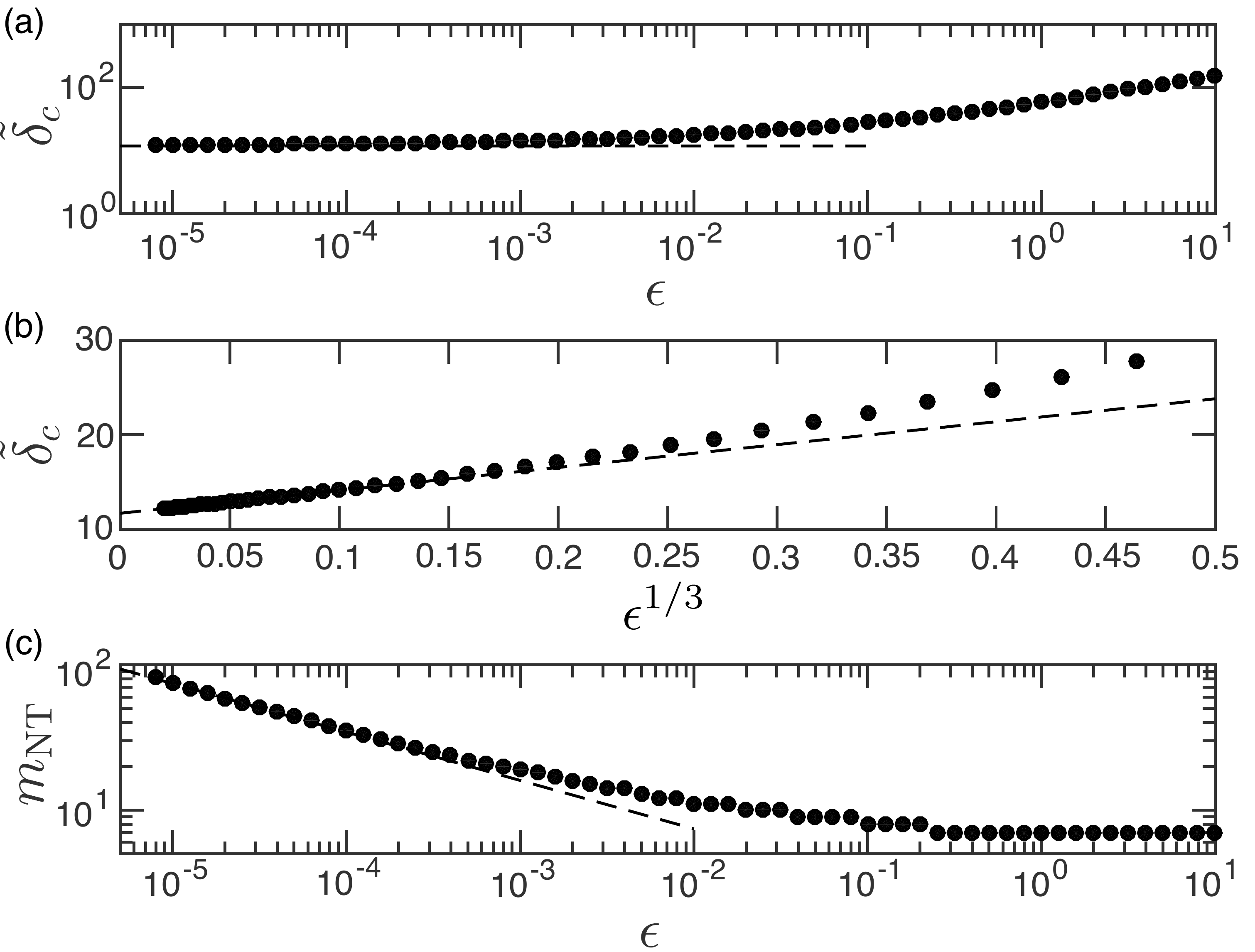}
\caption{Reanalysis of the numerical results of Box \emph{et al.}~\cite{Box17} for the onset of wrinkling  in an indented, floating elastic sheet of infinite radius; here their numerical results are presented in a way that highlights the scaling laws  for $\epsilon\ll1$ that are derived in the current paper. (a) The indentation depth at which wrinkles appear, $\twoc(\epsilon)$, tends to the expected membrane theory result \cite{Vella15}, $\awrink\approx11.75$ (dashed line). (b) Furthermore, the deviation from the membrane theory result scales according to \eqref{eqn:deltacNT}, with the dashed line showing the result $\twoc=\awrink+24.23\epsilon^{1/3}$ (with the pre-factor in this linear relationship determined by a best fit of those data points with the smallest values of $\epsilon$). (c) The wrinkle number at onset, $\mNT$, reproduces the scaling expected from \eqref{eqn:mNT} with the dashed line showing $\mNT\approx 1.6\epsilon^{-1/3}$ (with the pre-factor determined by a best fit of points with smallest $\epsilon$).}
\label{fig:NTnums}
\end{figure}

\subsection{Intermediate FT behavior}

Having understood how the onset of wrinkles depends on the finite size and finite bendability of the sheet, we now move on to study in more detail how the behavior of the system changes beyond the threshold. In particular, we focus on the limit of relatively large ($\Rnd\gg1$) and infinitely bendable  ($\epsilon\to0$) sheets for relatively large indentation depths, $\tdelta\gg\awrink$. We shall begin with the second Far-from-Threshold regime, FT-II, which develops with increasing indentation depth, and then move on to discuss how this regime morphs into the final FT regime, FT-III, in which wrinkles decorate the whole sheet, apart from a small inner tensile core.

\subsubsection{Asymptotic results in FT-II}

To obtain a quantitative description of the FT-II regime, we consider the limit of large indentation depth, $\tdelta\gg1$; here, it is possible to obtain asymptotic results for the wrinkle positions and indentation force. The detailed asymptotic arguments are given in Appendix \ref{Appendix:FT2}, but the key results are that wrinkles occupy the region $\tLi<\tr<\tLo$ with
\beq
\tLo\approx\frac{\Gamma(2/3)^{3/4}}{2\sqrt{ 3}\Gamma(1/3)^{3/4}}\frac{\tdelta^{3/2}}{(\log\tdelta)^{3/4}}\approx0.173\frac{\tdelta^{3/2}}{(\log\tdelta)^{3/4}},
\label{eqn:FT2LO}
\eeq and
\beq
\tLi\approx1.058\frac{\tdelta^{1/2}}{(\log\tdelta)^{5/4}}.
\label{eqn:FT2LI}
\eeq Furthermore, the dimensionless indentation force is given by
\beq
\tF\approx2.258 \frac{\tdelta^{2}}{(\log\tdelta)^{1/2}}.
\label{eqn:FT2F}
\eeq

We note that the primary scalings of \eqref{eqn:FT2LO}--\eqref{eqn:FT2F}, \emph{i.e.}~the exponents of $\tdelta$, are as anticipated in the scaling analysis of \S\ref{sec:scalingFT} --- compare, for example, \eqref{eqn:FT2LO} and \eqref{eqn:FT2LI} with (\ref{eq:FT-II-Scale}a) and (\ref{eq:FT-II-Scale}b), respectively. However, we note that in each case there is a logarithmic correction that could not have been anticipated through such a scaling analysis and, further, that the logarithmic terms themselves show power-law behavior. These rather exotic logarithmic corrections can be verified by comparison with suitable rescalings of our numerical results, as shown in fig.~\ref{fig:FT2nums}. (In particular note that to test a predicted scaling relation of the form $y\sim x^\alpha/(\log x)^\beta$ we present a log-plot of $yx^{-\alpha}$ versus $\log x$, which ought to yield a linear relationship with slope $-\beta$ on these logarithmic scales.) We see that the expected powers of $\log\tdelta$ are observed in each case, and that the associated pre-factors are consistent with the numerical results; however, these results involve the approximate inversion of transcendental expressions for $\Li$ and $\Lo$ so that some super-logarithmic corrections remain and lead to a noticeable disagreement, even when $\tdelta=O(10^8)$ (which is well beyond the experimentally accessible regime!).

\begin{figure}
\centering
\includegraphics[width=0.95\columnwidth]{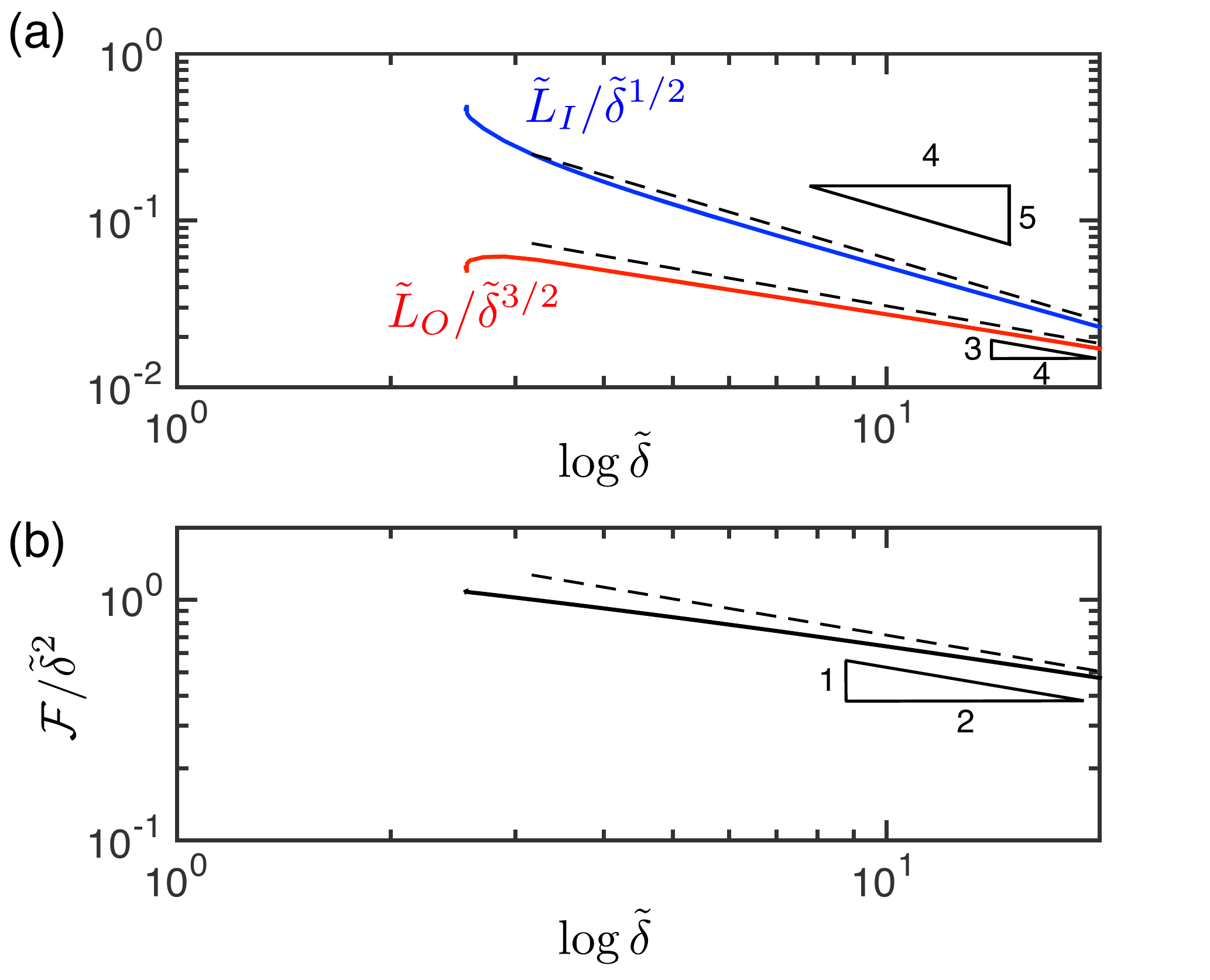}
\caption{ Numerical results (solid curves) replotted to interrogate the expected logarithmic corrections of \eqref{eqn:FT2LO}, \eqref{eqn:FT2LI} and \eqref{eqn:FT2F}, which are represented by the dashed lines in each case. (a) The wrinkle extents $\tLo\sim\tdelta^{3/2}/(\log\tdelta)^{3/4}$ and $\tLi\sim\tdelta^{1/2}/(\log\tdelta)^{5/4}$, as expected from \eqref{eqn:FT2LO} and \eqref{eqn:FT2LI}, respectively. (b) The indentation force $\tF\sim\tdelta^2/(\log\tdelta)^{1/2}$, as expected from \eqref{eqn:FT2F}. Note that in each case the $x$-axis uses a logarithmic scale for $\log\tdelta$ (results are shown for $\tdelta\lesssim \exp(20)\approx5\times10^8$). }
\label{fig:FT2nums}
\end{figure}

A central observation from the above asymptotics is that the outer position of the wrinkles $\tLo\sim\tdelta^{3/2}$, up to logarithmic factors. As a result, it is clear that wrinkles will grow quickly with indentation depth and, ultimately, must reach the edge of the sheet, \emph{i.e.}~$\tLo=\Rnd$. As noted earlier, this marks a dramatic change in the behavior of the system. We therefore turn to studying more precisely how wrinkles reach the film edge: the transition from FT-II to FT-III.

\subsubsection{Transitioning from FT-II: The role of finite size}

The asymptotic results above were presented for the case of an infinite sheet radius, $\Rnd=\infty$. To understand how the FT-II regime transitions to FT-III we must first understand how the finite size of the sheet affects the propagation of the outer edge of the wrinkles, $\tLo$, in regime FT-II. Numerical results that show the effect of finite $\Rnd$ are presented in fig.~\ref{fig:ForceLawNums}c (force) and fig.~\ref{fig:WrinkleExtents}b (wrinkle positions). We do not attempt to determine the asymptotic behavior of the wrinkle length $\tLo$ as wrinkles approach the edge of the sheet, \emph{i.e.}~$\tLo\to\Rnd$, though such an analysis may be possible. Instead, we shall only describe the value of the indentation depth $\tdelta$, denoted $\tdeltaastast$, at which wrinkles reach the edge. A simple (approximate) inversion of \eqref{eqn:FT2LO} with $\tLo=\Rnd$ and $\tdelta=\tdeltaastast$ yields
 \beq
\tdeltaastast\approx\frac{2^{7/6}\Gamma(1/3)^{1/2}}{3^{1/6}\Gamma(2/3)^{1/2}} \Rnd^{2/3} (\log\Rnd)^{1/2}\approx 2.63 \Rnd^{2/3} (\log\Rnd)^{1/2}.
\label{eqn:dcritFTIItransAsy}
 \eeq This result is plotted, together with the numerically determined dependence of $\tdeltaastast(\Rnd)$ in fig.~\ref{fig:RegimeDiag}. Note that for experimentally realizable values of the film radius, $\Rnd\lesssim10$, the critical value $\tdeltaastast\leq2\twoc$: wrinkles propagate to the edge of the sheet very quickly as the indentation depth increases beyond the wrinkling threshold $\twoc$.

\subsection{Regime FT-III ($\tdelta\gg\Rnd^{3/2}\gg1$)}


The final regime we consider, in which wrinkles have propagated to the edge of the sheet, FT-III, was studied in detail by Vella \emph{et al.} \cite{Vella15}. Here, we summarize their main results.

Once in regime FT-III, the inner edge of the wrinkle pattern rapidly progresses towards the indentation point; in particular, for $\tdelta\gg1$
\beq
\frac{\tLi}{\Rnd^{1/3}}\approx6.2003 ~\left(\tdelta/\Rnd^{2/3}\right)^{-2}.
\label{eqn:LiFT3}
\eeq The sheet thus becomes asymptotically covered in wrinkles while the indentation force becomes independent of the elastic properties of the sheet \cite{Vella15}:
\beq
\tF\approx4.58\Rnd^{2/3}\tdelta
\label{eqn:ForceFT3}
\eeq so that the large indentation stiffness
\beq
k=\frac{\tF}{\tdelta}\approx4.58\Rnd^{2/3}.
\label{eqn:FT3Stiffness}
\eeq This result is compared to the numerical results in fig.~\ref{fig:ForceLawNums}c.

\section{An energetic perspective \label{sec:discussion}}

We have shown that the indentation of a floating elastic sheet passes through a number of different regimes as the indentation depth $\delta$ increases. In this section we will address, at a qualitative level, the various energetic mechanisms through which the work done by the indenter is stored in this liquid-vapor-solid system. We will show that in each regime the energy is governed by a subset of these storage mechanisms and highlight  regime FT-III, in which the dominant energy components are gravitational potential energy of the displaced liquid and surface energy of the liquid--vapor interface. This type of energetic balance signals the emergence of an  ``asymptotically isometric mechanics", whereby only a negligible part of the indenter's work is transferred to the sheet, making it a bad capacitor of mechanical work \cite{Vella15}. Turning to the spatial distribution of the respective energy densities, we will show that even though the overall elastic energy due to straining the sheet becomes negligible, its density is prominent within a core zone around the indenter, signalling a novel type of energy (and stress) focusing.  
Probing further the nature of energy focusing in the unwrinkled core, we will highlight a potential route by which a wrinkled state of the type described in the preceding sections, may become energetically unstable.

\subsection{Energy storage mechanisms} 

We begin by discussing the various energies of the system (whose simultaneous minimization underlies the FvK equations and boundary conditions discussed in \S\ref{sec:detcalcs}). 
\\

$\bullet$ \underline{Liquid gravitational potential energy:} The gravitational potential energy (gpe) of the liquid displaced by the sheet, $\Ugrav$,  is easily estimated by considering an average vertical deflection, $\sim \delta$, over a horizontal area, $\sim \lcurv^2$. Hence: 
\begin{equation}
\Ugrav \sim \ksub \lcurv^2 \delta^2  \ \ ; \ \ \ksub = \rhol \cdot g
\label{eq:Ugrav}
\end{equation} 
\\    

$\bullet$ \underline{Liquid-vapor surface energy:}  The energetic cost of the liquid-vapor surface, which is exposed through the inward displacement of the sheet's edge, is 
$\Usurf = 2\glv\pi\Rfilm |\rmu_r(\Rfilm)|$ (note that in Ref.~\cite{Vella15}, $W_{\rm surf}  = - \Usurf$ was interpreted as the work done on the sheet by the liquid--vapor interfacial tension that pulls on its edge). Since we are interested only in the $\delta$-dependence of the energy, we evaluate the surface energy with respect to a ``base value", associated with the planar state ({\emph{i.e.}} $\tdelta=0$) of a floating sheet  of radius $\Rfilm$, and subject to a tension $\glv$ exerted at its perimeter.  This base energy is
\begin{equation}
\Ustrainp = \Rfilm^2\glv^2/2Y \ .
\label{eq:Uplan}
\end{equation} 
As the indentation depth, $\tdelta$, increases, we find the asymptotic relations: 
\begin{equation}
\frac{\Usurf}{\Ustrainp}  \sim \left\{
\begin{array}{cc}
 \approx 0  & \tdelta \ll 1\  \\
 (\tdelta/\Rnd^{2/3})^3 ,&  1\ll \tdelta \ll \Rnd^{2/3},   \\
   (\tdelta/\Rnd^{2/3})^2,  & \tdelta \gg \Rnd^{2/3}.
\end{array} \right.
\label{eq:Usurf}
\end{equation} Note that in the second line we have employed \eqref{eqn:FT2LO}, whereas the third line can be understood by drawing an analogy between the radial profile and the deflection of an inextensible string \cite{Vella15}.  
\\

$\bullet$ \underline{Strain energy:}  The elastic energy due to straining the sheet (in comparison to a planar state with free boundaries) is $\Ustrain  = \int \upd^2x~ \sigma_{ij}\varepsilon_{ij}$ \cite{Chopin08}. Once again, since we are interested only in the $\delta$-dependence of the energy, we evaluate it relative to the base state energy, 
$\Ustrainp$, given in \eqref{eq:Uplan}. We obtain: 
\begin{gather}
\Delta\Ustrain  \sim \left\{
\begin{array}{cc}
\!\!\!\!\!\!\!\!\!\!\!\!\!\!\!\!\!\!\!\!\!\!\!\!\!\!\!\!\!\!\!\!\!\!\!\!\!\!\!\!\!\!\!\!\!\!\!\!\!\!\!\!\!\!\!\!\ \glv \delta^2, & \tdelta \ll 1  \\
\!\!\!\!\!\!\!\!\!\!\!\!\!\!\!\!\!\!\!\!\!\!\!\!\!\!\!\!\!\!\!\! \sqrt{Y\cdot \ksub} \ \delta^3,  & 1\ll \tdelta \ll \Rnd^{2/3}   \\
 \log\left(\tdelta/\Rnd^{1/3}\right) \cdot \Ustrainp, & \tdelta \gg \Rnd^{2/3}     
\end{array} \right.
\label{eq:Ustrain}
\end{gather} 
A few aspects in the above equation are noteworthy. First, in the linear regime, $\tdelta \ll 1$, the strain energy reflects a slight increase of the sheet's area (by an amount $\propto \delta^2$ in comparison to the area of the floating sheet), while retaining a nearly uniform, isotropic stress, $\glv$, throughout. Barring the proportionality constant, $k_0(\epsilon)$, this type of linear response is akin to the response of a fluid membrane to poking. 

Second, the estimate of the strain energy in regime FT-II ($1\ll \tdelta \ll \Rnd^{2/3}$) is independent of the pre-existing tension $\glv$ in the floating sheet. Moreover, this expression is similar to an expression for the ``bare strain" energy,
\begin{equation}
\Ustrainb  \sim Y (\delta/\lcurv)^4 \lcurv^2, 
\label{eq:Ustrain-bare}
\end{equation} 
acquired by a naturally-flat sheet upon developing Gaussian curvature over an area $\sim \lcurv^2$ \cite{Hure12}. 
Although, at the scaling level, the effect of wrinkles does not seem to have a large effect on the level of strain in this regime (since, $\Ustrain \sim  \Ustrainb$), there is actually a dramatic suppression of the strain level. This underlies the shrinking size of the purely tensile ``core" upon increasing $\tdelta$ in comparison to the axisymmetric, unwrinkled deformation, as demonstrated in Fig.~\ref{fig:WrinkleExtents}a.  

Finally, the third line of Eq.~(\ref{eq:Ustrain}) reflects the dramatic effectiveness of wrinkles in relieving the strain imposed by the indenter: a comparison between Eqs.~(\ref{eq:Ustrain}) and~(\ref{eq:Ustrain-bare}) reveals that in the absence of wrinkles the strain energy would continue to grow  considerably with indentation depth ($\sim \delta^3$). However, the formation of wrinkles enables the actual strain to remain almost at the level of the floating, pre-indented state of the system, with only a weak, logarithmic dependence on $\delta$. Such a response is an example of ``asymptotically isometric" mechanics \cite{Vella15}, on which we will elaborate further below. 

One may also note that in regime FT-III, the strain energy in the unwrinkled core scales in the same way as the base value $\Ustrainp$: $Y\cdot(\delta/\lcurv)^4 L_I^2\sim\Ustrainp$. The logarithmic dependence on $\delta$ seen on the third row of \eqref{eq:Ustrain} is associated with the residual energy in the wrinkled zone, $\int_{L_I}^{\Rfilm} r\,\upd r \ (\srr^2/Y) \sim \int_{L_I}^{\Rfilm}~\upd r/r$. In fact, an alternative, energy-based approach to derive the scaling relation for $L_I$ (Eq.~\ref{eq:FT-III-Scale}) is to consider $L_I$ as a free variable, and minimize $\Ustrain$ by varying $\Li$. This energy-based approach to obtain the core size, $L_I$, will re-surface later in our discussion.   
\\

$\bullet$ \underline{Bending energy:} The bending energy of the sheet is, $\Ubend \sim B \int \upd^2x (\kappa_{rr} + \kappa_{\q\q})^2 \approx B \int \upd^2x \ \kappa_{\q\q}^2 $, since the radial curvature, $\kappa_{rr} \approx  \zeta''(r)$, is negligible in comparison to the azimuthal curvature that stems from highly wrinkled azimuthal  undulations. 
Using Eqs.~\eqref{eq:wrinkle-pattern} and \eqref{eq:slaving} we obtain: 
\begin{equation}
\Ubend \sim B  \int_{L_I}^{L_O} \ \upd r \ \lambda^{-2}(r) \,|\rmu_r(r)|     
\label{eq:Ubend}
\end{equation}  
We focus on regime FT-III (for which the wrinkle contribution is the largest), and evaluate this integral by  ignoring the small deviation of the radial displacement, $\rmu_r(r)$, from the asymptotically-isometric behavior that is obtained as $\tdelta\to\infty$. In particular, we take:
\beq
\rmu_r(r)=-\frac{\glv \Rfilm}{ Y}\times\frac{\tdelta^2\Rnd^{-4/3}}{2[\Ai(0)]^2}\int_0^{\tfrac{r}{\Rnd^{1/3}\lc}}[\Ai'(\xi)]^2~\upd\xi.
\label{eqn:AsyDispProfile}
\eeq The integral \eqref{eqn:AsyDispProfile} can be determined analytically, but the main result is that
\beq
\rmu_r(r)\approx\begin{cases}
-C_1\delta^2r/\lcurv^2,\quad r\ll\lcurv\\
-C_2\delta^2/\lcurv,\quad r\gg\lcurv\\
\end{cases}
\label{eqn:urApprox}
\eeq where $C_1=3^{2/3}\Gamma(2/3)^2/[2\Gamma(1/3)^2]\approx0.266$ and $C_2=\Gamma(2/3)/[3^{2/3}\Gamma(1/3)]\approx0.243$ are constants determined through the asymptotic behavior of \eqref{eqn:AsyDispProfile}. (The behavior of the numerically determined horizontal displacement $\rmu_r(r)$  is shown in fig.~\ref{fig:radialdisp} together with the asymptotic result \eqref{eqn:AsyDispProfile} and the approximate results \eqref{eqn:urApprox}.) Concerning the wavenumber, $m(r) = 2\pi r/\lambda(r)$, we employ Eqs.~\eqref{eqn:LambdaLaw} and \eqref{eqn:Keff}, noting that in the curved zone, $r<\lcurv$, the wavelength is curvature-dominated ({\emph{i.e.}}~$\keff \approx \kcurv \sim Y (\delta/\lcurv^2)^2 $), while in $\lcurv<r<\Rfilm$ it is substrate-dominated  ({\emph{i.e.}} $\keff \approx \ksub \sim \rhol g$). Substituting in Eq.~(\ref{eq:Ubend}), we  obtain: 
\begin{equation}
\Ubend  \sim \begin{cases}
   (B\ksub)^{1/2} \Rfilm \cdot \frac{\delta^2}{\lcurv} ,\quad \Rnd^{2/3}\ll \tdelta\ll\Rnd^{4/3} \  \\
 (BY)^{1/2} \cdot \frac{\delta^3}{\lcurv^2},\qquad \qquad \tdelta \gg\Rnd^{4/3}.   
\end{cases}
\label{eq:Ubend-2}
\end{equation}

\begin{figure}
\centering
\includegraphics[width=0.9\columnwidth]{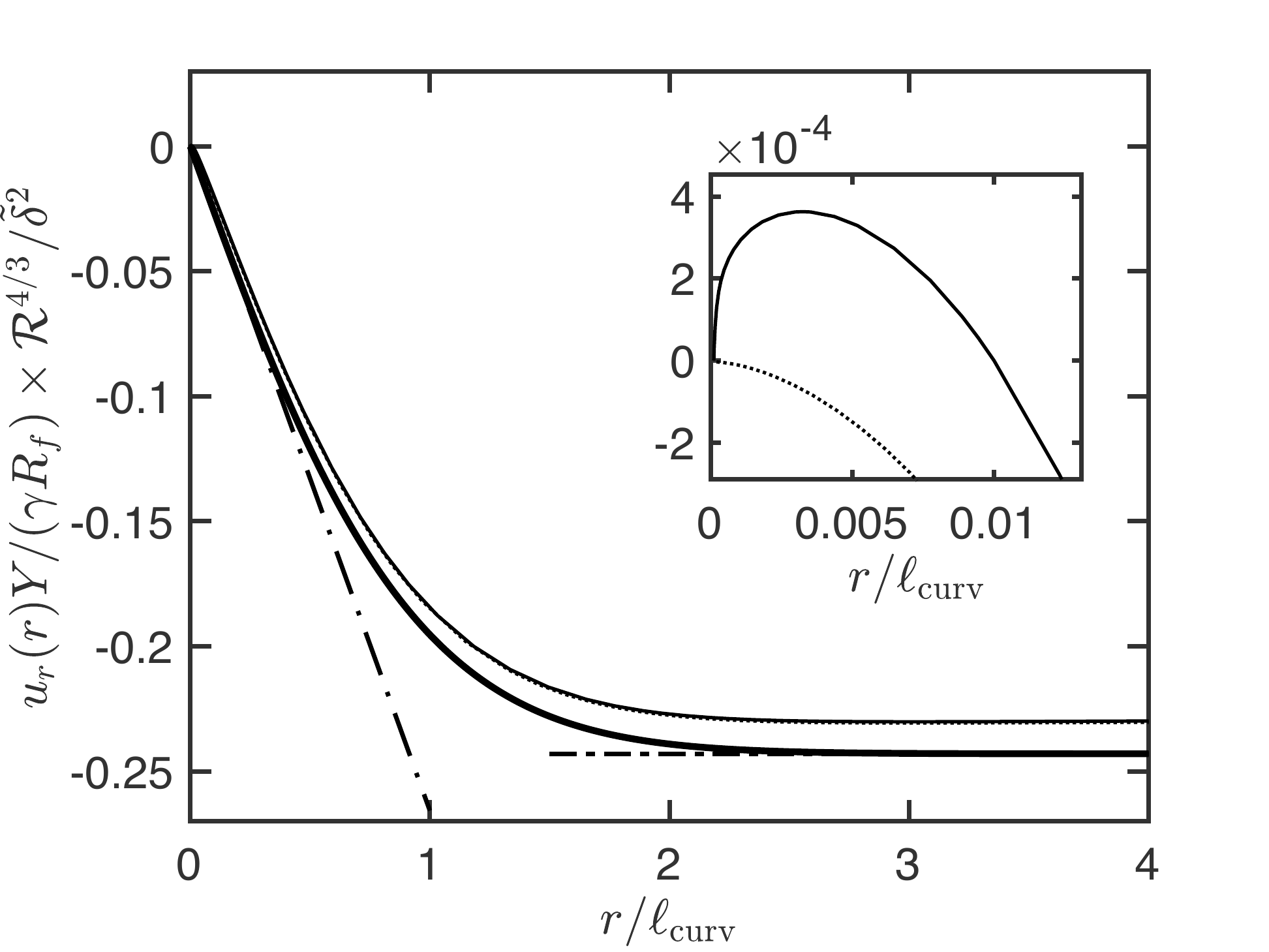}
\caption{The radial displacement, $\rmu_r(r)$, in regime FT-III. Numerical results are shown for the whole sheet when $\tdelta/\Rnd^{2/3}\approx25$ (corresponding to the inner tensile region being of size $\Li/\lcurv=0.01$) with Poisson ratio $\nu=0$ (thin solid black curve) and $\nu=1/3$ (dotted thin black curve). Main figure: The thick solid black curve shows the ``perfectly isometric" displacement profile obtained as $\tdelta\to\infty$, \eqref{eqn:AsyDispProfile}; this is in turn approximated by $\rmu_r \approx -C_1r \delta^2/\lcurv^2$ for $r\ll \lcurv$, and $\rmu_r \approx -C_2  \delta^2/\lcurv$ for $\lcurv\ll r<\Rfilm$ which are shown by the dash-dotted lines. (The values of the constants  $C_1$ and $C_2$ are given after \eqref{eqn:urApprox}.) Inset: Within the tensile region the radial displacement may be positive or negative, depending on the Poisson ratio. A zoom of the region $0<r\lesssim \Li$ shows that for $\nu=0$ (solid curve) $\rmu_r>0$ in the tensile zone but for $\nu=1/3$ (dotted curve) $\rmu_r<0$.}  
\label{fig:radialdisp}
\end{figure}

\subsection{Where does the indenter's work go?}

Our analysis in the preceding sections has focused on the radial profiles of the stress components, and the basic properties of the wrinkle pattern. With the aid of Eqs.~\eqref{eq:Ugrav}--\eqref{eq:Ustrain} and (\ref{eq:Ubend-2}), we will now show that,   
depending on the dimensionless depth, $\tdelta$, different combinations of the elastic energy of the sheet, gravitational potential energy of the displaced liquid, and surface energy of the liquid--vapor interface, are in balance.   
\\

\underline{Linear response:} In the first regime, which we identify as the ``true" linear response regime of the system ({\emph{i.e.}} valid for arbitrarily small values of $\tdelta \ll1$), the work done by the indenter, $ F\delta\approx k_0(\epsilon) \glv\delta^2$, 
goes into the gpe of the displaced liquid, and to a slight increase in the sheet's area (which increases by an amount $\sim \delta^2$ in comparison to the area of the floating sheet, while retaining a nearly uniform, isotropic stress, $\glv$, throughout): 
\begin{equation}
F\cdot \delta  \ \rightarrow  \  {\rm (liquid \ gpe)} \ + \ {\rm (additional \ sheet \ area)} 
\label{eq:response-1}
\end{equation}  
This type of linear response is akin to the response of a fluid membrane (though note that  the liquid--vapor surface is not changed in this limit). 
Here, the only indication that the sheet is made of a solid material is the weak, logarithmic dependence of the indentation stiffness, $k_0(\epsilon)$, in \eqref{eqn:K1asy}, 
on the bendability, $\epsilon^{-1}$. This ``true" linear response mechanism at arbitrarily small values of $\delta$, the constancy of the linear stiffness $F(\delta)/\delta$, underlies the plateaus shown in Fig.~\ref{fig:ForceLawNums}a for different values of $\epsilon$.    
\\

\underline{Geometrically nonlinear response:} 
As the indentation depth increases, the stress field in the sheet develops a highly anisotropic and non-uniform profile, signalling a different mechanism for the transfer of energy from the indenter. This does not have a parallel in the mechanics of indented fluid membranes, since there the stress remains isotropic and uniform, regardless of the deformed shape of the membrane. While a fraction of the indenter's work is conveyed to the gravitational potential energy (gpe) of displaced liquid (underlying the increase of $\lcurv$ with $\delta$), and a smaller fraction is conveyed to creating a new liquid--vapor surface area at the edge of the sheet,  the rest of the work is used to generate the corresponding non-uniform, anisotropic distribution of strain in the solid sheet, rather than to merely a net increase of area: 
\begin{eqnarray}
F\cdot \delta  \ \rightarrow  \  {\rm (liquid \ gpe)} \ &+ \  \text{ (new liquid--vapor area)} \nonumber \\
&+ \ {\rm (Hookean \ solid \ strain)}.
\label{eq:response-2}
\end{eqnarray} 
This is also reflected by the appearance of the stretching modulus, $Y$, and the substrate stiffness, $\ksub = \rhol g$, in the second line of Eq.~(\ref{eq:Ustrain}). 
Note that the above schematic expression differs in two ways from  the mechanics around the onset of wrinkling (and the corresponding NT regime), as well as regimes FT-I and FT-II. The first difference between the parameter regimes, $\tdelta \sim O(1)$ ({\emph{i.e.}}~at or slightly above threshold), and $1\ll\delta \ll \Rnd^{2/3}$ ({\emph{i.e.}}~wrinkles partially covering the sheet), is in the nature of the 
stress field that accommodates the indentation (purely tensile if the sheet is unwrinked or compression-free in the wrinkled case). The second difference is the emergence of the liquid--vapor interface as an effective energy storage mechanism (see Fig.~\ref{fig:energy-transition}). 

The nonlinear response that results from Eq.~(\ref{eq:response-2}), whereby the indentation stiffness, $F(\delta)/\delta$, increases with $\delta$, underlies the black curves in  Fig.~\ref{fig:ForceLawNums}a,b. 
\\

\underline{Asymptotically isometric, ``pseudo-linear" response:} 
Finally, when the indentation depth is sufficiently large, $\tdelta \gg \Rnd^{2/3}$, the wrinkles reach the edge (regime FT-III), and the energy balance undergoes yet another qualitative change, as can be revealed by comparing the energies $\Ustrain$ (\ref{eq:Ustrain}) and $\Ubend$ (\ref{eq:Ubend}) with the energies $\Ugrav$ and $\Usurf$ (Eqs.~\ref{eq:Ugrav} and \ref{eq:Usurf}, respectively).   
Now, the work done by the indenter is no longer stored by expanding or (inhomogenously) straining the sheet, but rather by increasing the interfacial energy of the liquid bath due to additional area that is uncovered at the edge of the sheet (last line of Eq.~\ref{eq:Usurf}). 
\begin{equation}
F\cdot \delta  \ \rightarrow  \  {\rm (liquid \ gpe)}  +  \text{ (new liquid--vapor  area)} 
\label{eq:response-3}
\end{equation} 
The response described by the above schematic expression, which underlies Eqs.~(\ref{eqn:ForceFT3},\ref{eqn:FT3Stiffness}), signifies the emergence of ``asymptotically isometric" mechanics \cite{Vella15}. To wit, let us note that since all but a negligible part of the indenter's work is stored in the liquid, the indentation stiffness, $k$, is independent of the elastic modulii ($Y$ and $B$) of the sheet. However, despite the explicit absence of solid energy from the RHS of (\ref{eq:response-3}), this expression does reflect the presence of a solid sheet that connects the indenter to the liquid--vapor interface. In fact, it is the dual nature of this solid sheet, being nearly inextensible on the one hand and highly bendable on the other hand, that determines the amount of liquid--vapor surface area that must be exposed upon increasing 
$\delta$, thus affecting an indentation stiffness, $F(\delta)/\delta = k$ (Eq.~\ref{eqn:FT3Stiffness}). Notably, while being independent of the elastic moduli of the solid, the indentation stiffness, $k \sim \Rnd^{1/3} k_0$, is much larger than its counterpart, $k_0$, in the ``truly linear" regime at sufficiently small $\delta$. The linear response that results from Eq.~(\ref{eq:response-3}), (which we refer to as ``pseudo-linear" since it does not extend to arbitrarily small indentation depth), underlies the plateau regions in fig.~\ref{fig:ForceLawNums}c where different colors correspond to different values of the ratio $\Rnd = \Rfilm/\lc$.

Taken together, Eqs.~\eqref{eq:response-1}--\eqref{eq:response-3} span a surprisingly rich mechanics, whose origin is purely geometric (since the material response of the solid is purely Hookean: the stress and strain in any small piece of the sheet remain proportional throughout). This is depicted schematically in Fig.~\ref{fig:energy-transition}, which exhibits the variation of the energies, $\Ugrav, \Usurf,\Ustrain,\Ubend$, upon increasing $\tdelta$.  
At sufficiently small $\tdelta$, we observe the expected linear response, with an indentation stiffness, $k_0$, that is determined (almost) solely by the pre-existing tension in the sheet; at sufficiently large $\tdelta$, the response is once again linear (in the sense that $F(\delta)/\delta$ is a constant $k \gg k_0$), but this ``pseudo-linear" response results from a highly nonlinear geometric constraint, imposed by the near-inextensibility and high-bendability of the sheet. The transition from the ``truly linear" response at small $\tdelta$ to the ``pseudo-linear" response at large $\tdelta$ (neither of which depends on the elastic constants of the sheet, except for logarithmically) is mediated by a (geometrically) nonlinear response (\ref{eq:response-2}). Furthermore, it is only in this transition region that the force is explicitly affected by the sheet's stretching modulus, $Y$. Such an evolution from one constant stiffness regime to another one, is similar to the behavior observed in the indentation of a pressurized elastic shell \cite{Vella12,Vella15EPL}.  

\begin{figure}
\centering
\includegraphics[width=0.9\columnwidth]{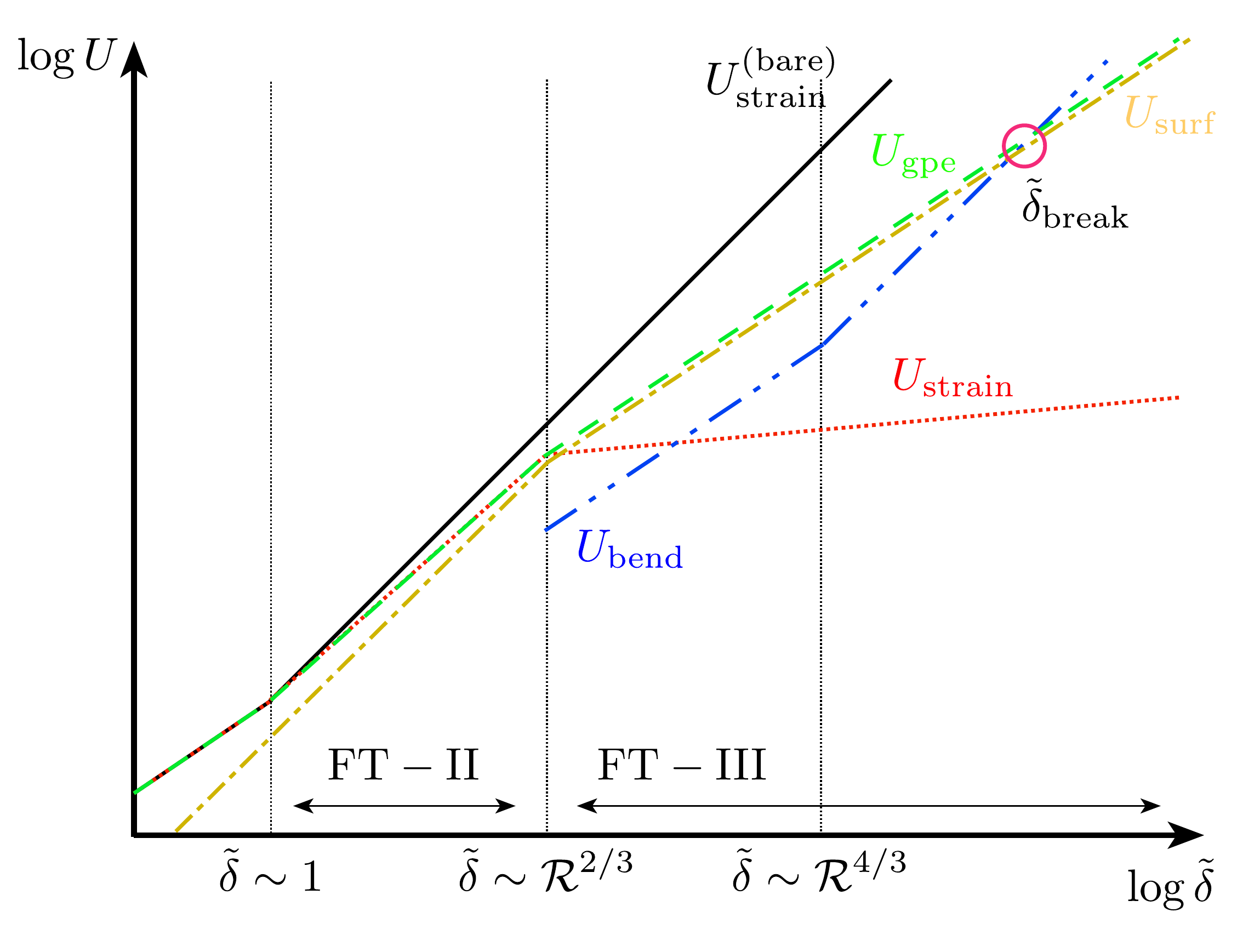}
\caption{(Color online) Schematic illustration of the evolution of the various energies in the problem with indentation depth $\tdelta$. In the absence of wrinkling, the strain energy of the system, $\Ustrainb$, increases according to \eqref{eq:Ustrain-bare} (black solid lines). The relaxation of the hoop stress through wrinkling allows the system to attain a lower energy state. Note that in regime FT-III the relaxed strain energy $\Ustrain$ (dotted red lines) becomes sub-dominant to the surface energy of uncovered liquid surface $\Usurf$ (dash-dotted yellow lines) and the gravitational potential energy of the displaced liquid $\Ugrav$ (dashed green lines). Note that the bending energy $\Ubend$ (blue dash-double dotted lines) is negligible in regime FT-II but  becomes larger than $\Ustrain$ if $\tdelta$ is sufficiently large within regime FT-III. At still larger $\tdelta\gtrsim\tdeltaTrans$ (the circled point) $\Ubend$ may become larger  even than $\Usurf$ and $\Ugrav$ so that we expect our analysis to break down. }
\label{fig:energy-transition}
\end{figure}

\subsection{Focusing of strain energy\label{sec:Focus}}

To further our understanding of the asymptotically isometric mechanics in the parameter regime FT-III, let us consider the spatial distribution of the energy density (per unit area). Figure~\ref{fig:energy-density} shows the  strain energy density in the sheet  ($\ustrain(r) = \tfrac{1}{2}\sigma_{ij}\varepsilon_{ij}$) and the density of the gpe of displaced liquid ($\ugrav(r) = \tfrac{1}{2} \ksub \zeta(r)^2$). Inspection of Fig.~\ref{fig:energy-density} reveals a remarkable feature of the asymptotically isometric response: while the explicit contribution of solid elasticity to the total energy of the system is negligible, the energy density is nevertheless dominated by $\ustrain$ within a small zone, $0<r\lesssim \Li$, which vanishes asymptotically as $\tdelta \to \infty$. One may easily confirm this energy focusing phenomenon, by noticing that in the unwrinkled core the strain energy is (approximately) 
$\ustrain \sim Y(\delta/\lcurv)^4$, whereas the small variation of the amplitude in this zone implies that $\ugrav \sim \ksub\delta^2$. Hence, in the unwrinkled core the ratio $\ustrain/\ugrav \sim (\tdelta/\Rnd^{2/3})^2 \gg 1$.  (Note that in the flat portion of the sheet, $r\gtrsim\lcurv$, $\ustrain/\ugrav\gg1$ again, simply because the vertical deflection decays exponentially in this portion of the sheet.)

\begin{figure}
\centering
\includegraphics[width=0.95\columnwidth]{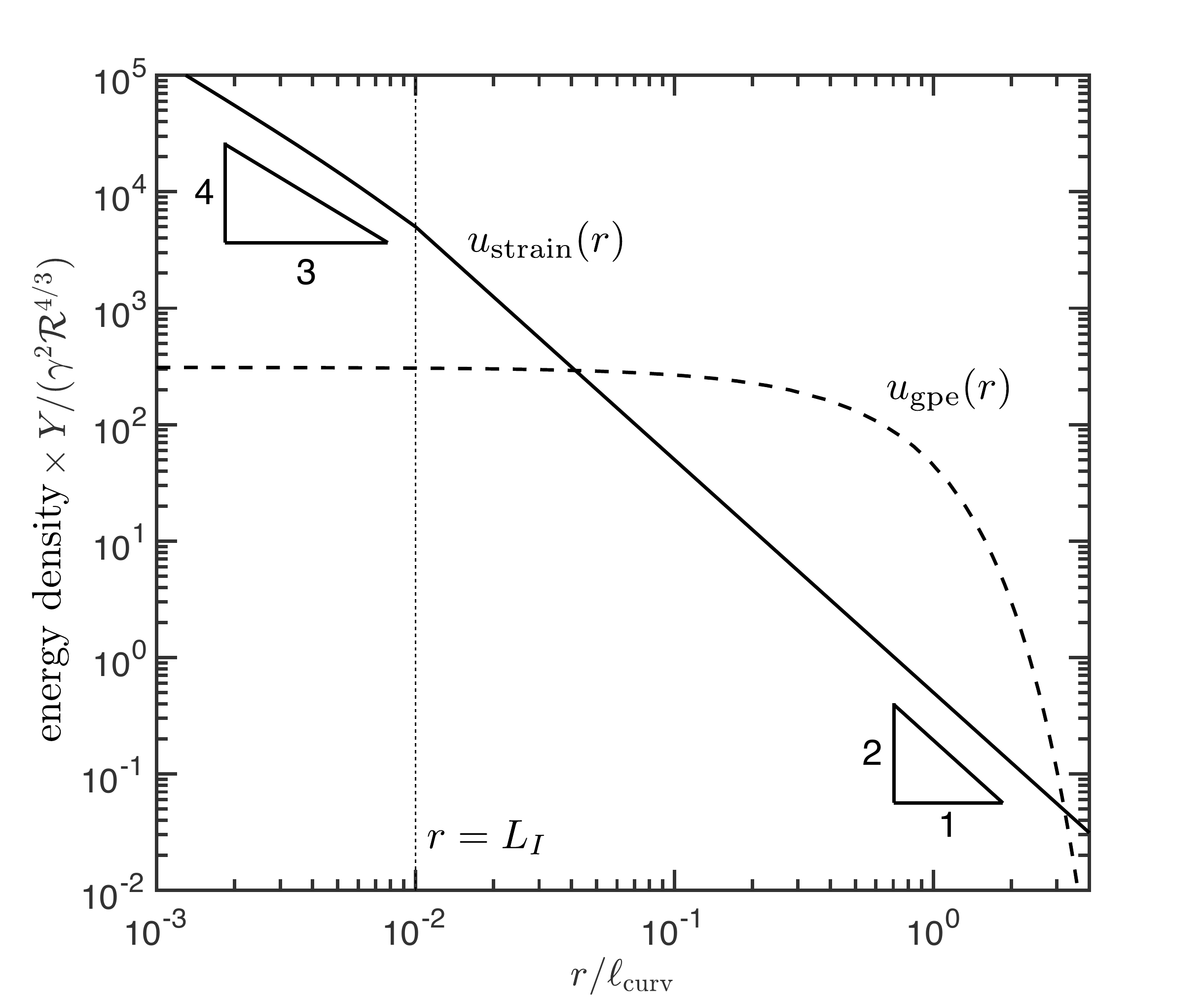}
\caption{The spatial distribution of the energy densities, $\ustrain(r)$ (solid curve) and $\ugrav(r)$ (dashed curve) in the parameter regime FT-III ({\emph{i.e.}}~with $\tdelta\gg\Rnd^{2/3}$). Here $\Li=10^{-2}\lcurv$ (as indicated by the vertical dotted line), which corresponds to $\tdelta/\Rnd^{2/3}\approx25$. Note that the elastic strain energy is focused in (and around) the unwrinkled core, $0<r\lesssim\Li$, where its density (which scales as $\delta^4$) exceeds the density of gravitational potential energy  (which scales as $\delta^2$ in $0<r<\lcurv$). The shrinking of the core upon increasing indentation depth ($L_I \sim \delta^{-2}$) underlies the negligibility of the total energy due to strain, $\Ustrain$, in comparison to the gravitational potential energy of the displaced liquid, $\Ugrav$, and liquid-vapor surface energy, $\Usurf$.}       
\label{fig:energy-density}
\end{figure}

Indeed, it is exactly this focusing of solid elastic energy (and stress) in the unwrinkled core that underlies the negligibility of $\Ustrain$, and hence the asymptotically isometric response. The energetic efficiency of this focusing mechanism is elucidated by considering the two contributions to $\Ustrain$: from the unwrinkled core we have $U_{\rm strain}^{\rm core} \sim Y (\delta/\lcurv)^4\cdot \Li^2$  (reflecting a strain $(\delta/\lcurv)^2$ within a disk of radius $\Li$) while from the wrinkled zone we have $U_{\rm strain}^{\rm wrinkle} \sim \glv \Rfilm^2 \log(\Rfilm/\Li)$ (reflecting a radial stress, $\srr = \glv\Rfilm/r$). Notice that the first term favors small $\Li$, whereas the second one favors large $L_I$;  minimization of their sum yields precisely the scaling law \eqref{eq:FT-III-Scale} for $\Li$, and the strain energy (\ref{eq:Ustrain}). Importantly,  $U_{\rm strain}^{\rm core}\sim \Ustrainp$, indicating that this mechanism focuses the strain of the pre-indented sheet in an ever shrinking core, with only a logarithmic amplification as $\tdelta$ increases. 

It is thus important to realize that while the explicit contribution of $\Ustrain$ to the total energy of the system in the parameter regime FT-III is negligible, the high energetic cost of straining the sheet underlies the wrinkle pattern and the associated mechanical response: the extent of the inner tensile zone is such that the strain energy is minimized and so even small deviations of the radial profile from the shape \eqref{eqn:MainTextAiry}, or $\Li$ from the scaling \eqref{eq:FT-III-Scale}, will considerably increase the strain energy towards its bare value, $\Ustrainb$ (Eq.~\ref{eq:Ustrain-bare}), at substantial energetic cost. Hence  ``macro-scale" features of the deformation (such as the importance of the horizontal scale $\lcurv$, and the size of the unwrinkled core $\Li$) are protected against small perturbations to the pattern, explaining their apparent robustness in indentation experiments (see, for example, Fig.~3b of \cite{Vella15} and Fig.~2F of \cite{Paulsen16}).       

\subsection{The energetic cost of wrinkling} 

An inherent assumption of tension field theory, and the consequent FT analysis, is the negligibility of the bending cost, 
$\Ubend$: our analysis is essentially an asymptotic expansion of the FvK equations around the singular limit of infinite bendability, $\epsilon=0$, assumed by tension field theory with the explicit cost of wrinkles compared to a comparable cost of deforming an ``effective substrate" \cite{Cerda03,Paulsen16}. Hence, our analysis hinges on the condition,
\begin{equation}
\Ubend \ll F\cdot \delta \ , 
\label{eq:cond-FT}
\end{equation}
which expresses that the indenter's work, $F\cdot\delta$, is stored in other, ``dominant", mechanisms thanks to the cheap cost of bending. In regimes FT-II and FT-III these mechanisms are described schematically in Eqs.~(\ref{eq:response-2}) and (\ref{eq:response-3}), respectively. 

One can verify (through some tedious calculation) that condition (\ref{eq:cond-FT}) is well satisfied in regime FT-II (as well as for smaller values of $\tdelta$), as long as the sheet is highly bendable, \emph{i.e.}~provided $\epsilon \ll 1$. However, regime FT-III introduces some subtleties that may have physical implications, as we elaborate below. 

Consider first the sub-regime of FT-III, $\Rnd^{2/3} \ll \tdelta \ll \Rnd^{4/3}$. Here, the bending energy is controlled by the cost of wrinkles in the flat part of the sheet, $\lcurv <r<\Rfilm$, as  indicated by the first line of Eq.~\eqref{eq:Ubend-2}. Comparing $\Ubend$ to $\Ugrav$ (\ref{eq:Ugrav}), which is the source of the dominant energy (along with $\Usurf$), shows that the inequality (\ref{eq:cond-FT}) is satisfied for any $\epsilon\ll1$. 

Turning now to the second sub-regime of FT-III, $\tdelta \gg \Rnd^{4/3}$, we note that $\Ubend \sim \tdelta^3$, whereas the dominant energies ($\Ugrav$ and $\Usurf$) scale as $\tdelta^2$. Hence, there exists $\tdeltaTrans$, such that for $\tdelta > \tdeltaTrans$, the energetic cost of wrinkles exceeds the dominant energy. Comparing  $\Ugrav$ with the relevant $\Ubend$ from the second line of Eq.~\eqref{eq:Ubend-2}, we find that
\begin{equation}
\tdeltaTrans \sim \Rnd^{4/3} \epsilon^{-1/2} \ . 
\label{eq:delta-up}
\end{equation}

We expect that as the indentation depth is increased, \emph{i.e.}~as $\tdelta \to \tdeltaTrans$, the wrinkle pattern may become energetically unstable and give way to another type of deformation that reflects the high energetic cost of straining the sheet, whilst also accounting for the bending energy. 

A natural candidate for such an instability might be the wrinkle-to-fold transition reported first by Holmes \& Crosby \cite{Holmes10}, in which wrinkles give way to localized folds (each of which accommodates a finite fraction of the excess length that had previously been stored in  wrinkles). However, the critical value at which this dramatic instability occurs is  $\tdelta=O(10-100)$ \cite{Holmes10,Paulsen16}, while $\tdeltaTrans=O(10^3)$ or higher for the same ultra-thin Polystyrene films \cite{Holmes10,Vella15,Paulsen16,Ripp18}, assuming an $O(1)$ pre-factor  in the scaling relation \eqref{eq:delta-up}. In addition to the different order of magnitudes between the observed critical value of the wrinkle-to-fold transition and the value of $\tdeltaTrans$, the strong thickness dependence of the latter (which corresponds to a physical indentation depth $\delta_{\mathrm{break}} \sim t^{-2}$) is in sharp contrast with that observed in the wrinkle-to-fold transition. Indeed, the observations in \cite{Holmes10}, as well as more recent experiments \cite{Paulsen16}, seem to suggest that folds appear at an indentation depth, $\delta_{\mathrm{fold}}$, that varies  weakly with thickness, apparently approaching  a constant value, that depends only on the capillary length $\lc$ and $\Rfilm$ as $t\to0$. Such an observation may indicate that the wrinkle-to-fold instability in indented sheets is not related to the increased bending cost, but is rather a ``purely geometric" instability, similar to wrinkle-to-fold transitions found in floating annular sheets \cite{Paulsen17}.   
\\
 
A careful reader may wonder whether the validity of tension field theory, and our resulting FT analysis of the wrinkled state, hinges on a stricter inequality than (\ref{eq:cond-FT}), \emph{e.g.}~$\Ubend \ll \Ustrain$. This latter condition is obviously satisfied in regime FT-II, where $\Ustrain$ is a dominant energy (accounting for a finite fraction of the work, $F\cdot \delta$). However, this is  not the case throughout the asymptotically isometric regime FT-III, where $\Ustrain$  is also a sub-dominant contribution to the total energy (see fig.~\ref{fig:energy-transition}).   To address this question, one must note that the reduction of the strain energy from the ``bare" value, $\Ustrainb$ (black curve in fig.~\ref{fig:energy-transition}), associated with imposing a Gaussian curvature on a naturally-planar sheet, to the ``residual" value, $\Ustrain$ (red dotted curve in fig.~\ref{fig:energy-transition}), requires the formation of a low-cost wrinkle pattern: it is the presence of wrinkles that allows the relaxation of the compressive hoop stress (by absorbing the consequent excess length of latitudes into wrinkly undulations), that allows the system to avoid radial strain. Hence, even if $\Ubend$ (\ref{eq:Ubend-2}) happens to exceed the residual strain energy, $\Ustrain$ (\ref{eq:Ustrain}), (as shown in fig.~\ref{fig:energy-transition}) the wrinkle pattern remains energetically favorable, as long as the weaker inequality~\eqref{eq:cond-FT} is satisfied.

There is, however, a subtle aspect of the energetic cost of wrinkles that pertains to the spatial distribution of the energy density and which may affect --- at least from a theoretical perspective --- the smoothness of the wrinkle pattern (\ref{eq:wrinkle-pattern}). To wit, let us note that the estimate (\ref{eq:Ubend-2}) of the bending energy ignores a putative divergence of the energetic cost of wrinkles in the vicinity of $r\approx \Li$ (due to a non-integrable divergence of the ``tensional stiffness", $\ktens (r) \sim \srr(\Li)(r-\Li)^{-1}$ \cite{Paulsen16}). Regularization of this divergence leads to the formation of a  ``boundary layer" --- a stress-focusing annulus around $r\approx \Li$ through which the amplitude of wrinkles decays into the tensile core \cite{Davidovitch12,Taylor15}, though other types of regularization have been proposed \cite{Bella14}.  
If $\Li \sim \tdelta^{-2}$ is sufficiently small, this boundary layer may interfere with the axisymmetric stress field in the core ($0<r<L_I$), signaling the possible emergence of an alternative stress-focusing mechanism to that described in \S\ref{sec:Focus}. This may in turn break the axial symmetry of the stress field in the vicinity of the wrinkle's tip and may also be a plausible mechanism underlying the formation of crumples \cite{King12}. Preliminary estimates suggest, however, that such a scenario occurs for $\tdelta$ larger than some critical value, whose scaling (with $\epsilon$ and $\Rnd$) is similar to $\tdeltaTrans$ (Eq.~\ref{eq:delta-up}). Such a mechanism seems also  to be of limited relevance to experiments in ultra-thin floating sheets: the underlying mechanism behind the breakdown of the `simple' wrinkled state described in this paper to crumpled and folded states remains poorly understood.

\acknowledgments

We thank the Aspen Center for Physics and the Parsegians' Casa F\'{i}sica for hospitality during this work. This research was supported by a Leverhulme Trust Research Fellowship (DV), the Zilkha Trust, Lincoln College (DV),  the European Research Council under the European Horizon 2020 Programme, ERC Grant Agreement no. 637334 (DV) and  NSF- CAREER Grant No.~DMR 11-51780 (BD). We have benefited from many discussions with M.~Adda-Bedia, E.~Cerda and N.~Menon; finally, we are grateful to M.~M.~Ripp and J.~D.~Paulsen for their willingness to share experimental results prior to publication and to D.~O'Kiely for comments on an earlier version of this paper.

\appendix

\section{ Detailed analysis of the three-region problem \label{Appendix:FT2}}

In this Appendix, we present a detailed analysis of the tension field theory model for the case in which wrinkles cover an annulus $\tLi<\tr<\tLo\leq\Rnd$. We shall keep this analysis as general as possible so that  a single analysis is able to cover the cases of FT-II and FT-III, as well as the transition between them. We shall assume that there are three regions: an inner tensile core ($0<\tr<\tLi$), a wrinkled annulus ($\tLi<\tr<\tLo$) and a tensile outer region ($\tLo<\tr<\Rnd$).

\subsection{The wrinkled zone and its exterior}

In the wrinkled annulus, the stress state is $\tsrr=C/\tr$, $\tsqq\approx0$ and so the vertical force balance equation may be written as
\beq
\frac{\upd^2\tzeta}{\upd \tr^2}=\frac{\tr}{C}\tzeta,
\eeq which has solution
\beq
\tzeta(\tr)=\alpha \Ai(\tr/C^{1/3})
\label{eqn:AppAiry}
\eeq for some constant $\alpha$. (Note that here we have neglected the second Airy stress function $\mathrm{Bi}(x)$, whose pre-factor we expect to be exponentially small in $\Rnd$.)

Eqn \eqref{eqn:AppAiry} shows that the vertical deflection of the sheet is expected to decay exponentially with horizontal distance from the indenter, over the length scale $C^{1/3}\lc$. As a result, we assume that the outer tensile region is approximately planar, which, from \eqref{eqn:FvK2ND}, suggests that the stress function in this outer region is of the form $\tpsi=\beta_1\tr+\beta_2/\tr$. Using the boundary conditions that $\tsrr=1$ at $\tr=\Rnd$ and $\tsqq=0$ at $\tr=\tLo$ we have that
\beq
\tpsi(\tr)=\frac{\tr+\tLo^2/\tr}{1+\tLo^2/\Rnd^2}.
\eeq This result also gives a relationship between the constant $C$ and the outer extent of the wrinkled region, since $C=\tLo\tsrr(\tLo)=\tpsi(\tLo)$. We therefore have that
\beq
C=\frac{2\tLo}{1+\tLo^2/\Rnd^2}.
\label{eqn:AppendC}
\eeq

\noindent We note at this point that there are two limits of \eqref{eqn:AppendC} that are of particular interest: (a) $\tLo\ll\Rnd$ (an effectively infinite sheet) for which $C\approx 2\tLo$ and (b) $\tLo=\Rnd$ (the wrinkles have reached the edge of the sheet) for which $C=\tLo=\Rnd$.

\subsection{The inner tensile region}

Having effectively solved the problem for the wrinkled annulus and the outer tensile region, we now turn to the inner tensile problem. In this region, $0<\tr<\Li$, we generalize the approach of Vella \emph{et al.} \cite{Vella15} and rescale the FvK equations \eqref{eqn:FvK1ND} and \eqref{eqn:FvK2ND} by letting
\beq
\rho=\tr/\Li,\quad \Psi(\rho)=\tpsi(\tr)/C, \quad Z(\rho)=\tzeta(\tr)/(C\tLi)^{1/2}.
\eeq The vertical force balance on the sheet is thus
\beq
\frac{1}{\rho}\frac{\upd}{\upd\rho}\left(\Psi\frac{\upd Z}{\upd \rho}\right)=\lambda^3Z+\frac{\tF}{2\pi} \tLi^{1/2}C^{-3/2}\frac{\delta(\rho)}{\rho}
\label{eqn:FvK1InnerTens}
\eeq while the compatibility of strains gives
\beq
\rho\frac{\upd }{\upd \rho}\left[\frac{1}{\rho}\frac{\upd }{\upd \rho}\left(\rho\Psi\right)\right]=-\frac{1}{2}\left(\frac{\upd Z}{\upd \rho}\right)^2,
\label{eqn:FvK2InnerTens}
\eeq where
\beq
\lambda=\tLi/C^{1/3}
\eeq is an unknown parameter that is to be determined as part of the solution.

The system of equations \eqref{eqn:FvK1InnerTens}--\eqref{eqn:FvK2InnerTens} is to be solved subject to the boundary conditions corresponding to  imposed (rescaled) displacements at the origin
\beq
Z(0)=-\frac{\tdelta}{(C\tLi)^{1/2}},\quad \lim_{\rho\to0}\left[\rho\Psi'-\nu\Psi\right]=0,
\label{eqn:InnerTensBC1}
\eeq and matching conditions on the stress components at the outer edge of the tensile region, i.e.
\beq
\Psi(1)=1,\quad \Psi'(1)=0.
\label{eqn:InnerTensBC2}
\eeq A final condition emerges from the requirement that the vertical displacement and slope of the sheet at the outer edge of the tensile zone must match smoothly to that at the inner edge of the wrinkled zone, \eqref{eqn:AppAiry}. This leads to the requirement
\beq
\frac{Z(1)}{Z'(1)}=\frac{\Ai(\lambda)}{\lambda\Ai'(\lambda)}.
\label{eqn:InnerTensBC3}
\eeq

As it stands, the fourth-order system \eqref{eqn:FvK1InnerTens}--\eqref{eqn:FvK2InnerTens} with the five boundary conditions  \eqref{eqn:InnerTensBC1}--\eqref{eqn:InnerTensBC3} is an over-determined system; this can readily be solved numerically to determine the rescaled inner wrinkle position in terms of the rescaled indentation depth, \emph{i.e.}~$\lambda=f(\tdelta/(C\tLi)^{1/2})$ with  $f(x)$ a numerically determined function. To `undo' the rescaling, we require one further condition. The appropriate condition differs in  regimes FT-II and FT-III, and so we consider these regimes separately. However, first we shall consider the limit of small tensile regions, $\lambda=\tLi/C^{1/3}\ll1$, which, as we shall see, is relevant for extremely large indentation depths, $\log\tdelta\gg1$, and hence is relevant to both regimes FT-II and FT-III.

In the limit $\lambda\ll1$, \eqref{eqn:FvK1InnerTens} may be approximated by 
\beq
\frac{1}{\rho}\frac{\upd}{\upd\rho}\left(\Psi\frac{\upd Z}{\upd \rho}\right)=\frac{\tF}{2\pi} \tLi^{1/2}C^{-3/2}\frac{\delta(\rho)}{\rho},
\eeq which may be integrated once to give
\beq
\Psi\frac{\upd Z}{\upd \rho}=\frac{\tF}{2\pi} \tLi^{1/2}C^{-3/2}.
\label{eqn:SmallLamFirstInt}
\eeq Eqn \eqref{eqn:SmallLamFirstInt} may then be used to eliminate $\upd Z/\upd\rho$ in the second FvK eqn \eqref{eqn:FvK2InnerTens} in favour of $\Psi$. The resulting equation may be solved analytically \cite{Chopin08,Vella15,Vella17}, albeit implicitly, to give
\beq
Z(\Phi)=-\frac{\tdelta}{(C\tLi)^{1/2}}+\frac{2}{A^{1/2}}\sinh^{-1}\bigl[(A\Phi)^{1/2}\bigr],
\label{eqn:InnerTensDelta}
\eeq
\begin{eqnarray*}
\frac{A}{2(1+A)^{1/2}}&(1-\rho^2)=(1+A)^{1/2}-\Phi^{1/2}(1+A\Phi)^{1/2}\\
&+A^{-1/2}\left\{\sinh^{-1}\bigl[(A\Phi)^{1/2}\bigr]-\sinh^{-1}(A^{1/2})\right\}
\end{eqnarray*} and
\beq
\Psi=\Phi/\rho.
\eeq Here $A\approx-0.697$ is a constant related to the dimensionless force via $\tF\Li^{1/2}/\bigl[2\pi C^{3/2}\bigr]=(1+A)^{-1/2}$ and is determined by the solution of
\beq
\frac{A+2}{2(1+A)^{1/2}}=A^{-1/2}\sinh^{-1}\bigl(A^{1/2}\bigr).
\eeq We therefore have immediately that
\beq
\tF=\frac{2\pi}{\sqrt{1+A}}C^{3/2}\tLi^{-1/2}\approx11.4\,C^{3/2}\tLi^{-1/2}.
\label{eqn:InnerTensForce}
\eeq 

Our main interest at this point lies in the determination of the parameter $\lambda$ as a function of $\tdelta$. We find from \eqref{eqn:InnerTensBC3} that
\beq
\frac{2\pi\lambda Z(1)C^{3/2}}{\tF\tLi^{1/2}}=\frac{\Ai(\lambda)}{\Ai'(\lambda)}\to-\frac{\Gamma(1/3)}{3^{1/3}\Gamma(2/3)}.
\eeq  The value of $Z(1)$ here can be related to the indentation depth $\tdelta$ via \eqref{eqn:InnerTensDelta}. We note that the second term on the RHS of \eqref{eqn:InnerTensDelta} is an $O(1)$ constant and hence can be neglected in comparison to $\tdelta/(C\tLi)^{1/2}\gg1$ for sufficiently large $\tdelta$; making this approximation we find that
\beq
\tLi \approx\left(\frac{\Gamma(1/3)}{3^{1/3}\Gamma(2/3)\sqrt{1+A}}\right)^2C^{5/3} \tdelta^{-2}\approx 6.20C^{5/3} \tdelta^{-2}.
\label{eqn:InnerTensLiDet}
\eeq

\noindent We therefore have three equations, \eqref{eqn:AppendC}, \eqref{eqn:InnerTensForce} and \eqref{eqn:InnerTensLiDet} relating the four unknowns ($C$, $\tF$, $\tLi$ and $\tLo$). To progress further, we must consider more specifically which regime we lie in. 

\subsection{FT-III}

Though it occurs for larger indentation depths, the case of FT-III is actually simpler to understand than regime FT-II; this is because, by definition, $\tLo=\Rnd$ in this regime, and hence $C=\Rnd$. Eqn \eqref{eqn:InnerTensLiDet} immediately gives \eqref{eqn:LiFT3}, which is precisely the result reported previously \cite{Vella15}. Note that in this case, it is clear that the asymptotic requirement $\lambda\ll1$ holds for $\tdelta\gg1$ since $\lambda=\tLi/C^{1/3}\sim\Rnd^{4/3}\tdelta^{-2}\ll1$ provided that $\tdelta\gg\Rnd^{2/3}$. 

For completeness, we also note that the indentation force is given by substituting \eqref{eqn:LiFT3} and $C=\Rnd$ into \eqref{eqn:InnerTensForce},  leading immediately to \eqref{eqn:ForceFT3}.

\subsection{FT-II}

In the case where the wrinkles have not yet reached the edge, the coefficient $C$ is not immediately determined, and so an alternative closure condition is required. In this case, the fact that there are two tensile regions ($0<\tr<\tLi$ and $\tLo<\tr<\Rnd$), together with the continuity of the stresses, membrane slope and deflection across  the interfaces $\tr=\tLi$ and $\tr=\tLo$, ensures  that the horizontal displacement at the inner and outer edges of the wrinkle pattern should match, \emph{i.e.}~$u(\Li^-)=u(\Lo^+)$ (more details of this argument are given elsewhere \cite{Vella15,Vella15EPL}). Using the expression for geometrically nonlinear strains we then find that
\begin{eqnarray*}
C\log(\Lo/\Li)&=&\tfrac{1}{2}\int_{\tLi}^{\tLo}\left(\frac{\partial\tzeta}{\partial \tr}\right)^2~\upd \tr\\
&=&\frac{\alpha^2}{2C^{1/3}}\int_{\lambda}^{\lambda \Lo/\Li}\left[\Ai'(\xi)\right]^2~\upd \xi.
\end{eqnarray*}

Since we expect $\lambda\ll1$ and $\Lo/\Li\gg1$, it is natural to assume that the upper and lower limits in the integral may be replaced by $0$ and $\infty$, respectively. Further, we assume that $\alpha\approx\tdelta/\Ai(0)$.  We shall make these assumptions for now and verify subsequently that the error introduced by this assumption is at the same order as that already made elsewhere in the calculation. We find that
\beq
C^{4/3}\log(\Lo/\Li)\approx\frac{\Gamma(2/3)^2}{2\cdot 3^{1/6}\pi}\tdelta^2.
\label{eqn:FT2Close}
\eeq

Examining the two equations \eqref{eqn:InnerTensLiDet} and \eqref{eqn:FT2Close}, we see that we have two equations in three unknowns ($\tLi$, $\tLo$ and $C$). The final equation is the expression for $C(\tLo,\Rnd)$ given by \eqref{eqn:AppendC}. This is effectively  a quadratic equation for $\tLo/\Rnd$ as a function of $C$, which may then readily be inverted and applied throughout FT-II. However, to make analytical progress, we  first consider the limit of FT-II in an infinite sheet: $\tLo/\Rnd\ll1$ and $C\approx2\tLo$, as already discussed. We therefore must solve:
\beq
\tLi\approx 6.20 (2\tLo)^{5/3}\tdelta^{-2},
\eeq and
\beq
\tLo^{4/3}\log(\Lo/\Li)\approx\frac{\Gamma(2/3)^2}{2^{7/3}\cdot 3^{1/6}\pi}\tdelta^2
\eeq simultaneously. The leading order results of this inversion are given in the main text as eqns \eqref{eqn:FT2LO} and \eqref{eqn:FT2LI}. Substituting these results into \eqref{eqn:InnerTensForce} gives the force law \eqref{eqn:FT2F}.

For more general values of $\tLo/\Rnd$, we do not proceed asymptotically and merely note that the aspect of most practical interest is the value of $\tdelta$ at which wrinkles reach the outer edge of the sheet, \emph{i.e.}~$\tLo=\Rnd$. We denote this critical indentation depth by $\tdeltaastast$ and note that there $\tLo=C=\Rnd$. It is then straightforward to use \eqref{eqn:FT2LO} to show that $\Rnd\approx0.173\tdeltaastast^{3/2}/(\log\tdeltaastast)^{3/4}$ with $\tLo=\Rnd$, which has approximate solution \eqref{eqn:dcritFTIItransAsy}.

Finally, we check when the various simplifying assumptions hold. First, regarding the assumption that  $\lambda\ll1$, recall that $\lambda=\tLi/C^{1/3}\sim\tLi/\tLo^{1/3}\sim(\log\tdelta)^{-1}$, where we have used \eqref{eqn:FT2LO} and \eqref{eqn:FT2LI}. Hence, we see that the requirement that $\lambda\ll1$ \emph{is} satisfied in the limit $\log\tdelta\gg1$ but that this requires $\tdelta\ggg1$. This explains the relatively slow convergence of the numerics to the scalings obtained for $\tLi, \tLo$ and $\tF$. Second, we assumed that $\int_{\lambda}^{\lambda \Lo/\Li}\left[\Ai'(\xi)\right]^2~\upd \xi\approx\int_{0}^{\infty}\left[\Ai'(\xi)\right]^2~\upd \xi$, and that $\alpha\approx\tdelta/\Ai(0)$. Both of these approximations are correct to leading order in $\lambda$ and hence are valid whenever $\log\tdelta\gg1$; this is consistent with the assumption $\tdelta/(C\tLi)^{1/2}\gg1$ made in deriving \eqref{eqn:InnerTensLiDet}.


\begin{thebibliography}{10}


\bibitem{HernandoPerez12}
M.~Hernando-P\'{e}rez, R.~Miranda, M.~Aznar, J.~L. Carrascosa, I.~A.~T. Schaap,
  D.~Reguera, and P.~J. de~Pablo,
\newblock Small {\bf 8}, 2366 (2012).

\bibitem{arnoldi00}
M.~Arnoldi, M.~Fritz, E.~Ba\"{u}erlein, M.~Radmacher, E.~Sackmann, and
  A.~Boulbitch,
\newblock Phys. Rev. E {\bf 62}, 1034 (2000).

\bibitem{Milani13}
P.~Milani, S.~A. Braybrook, and A.~Boudoaud,
\newblock J. Exp. Bot. {\bf 64}, 4651 (2013).

\bibitem{arfsten10}
J.~Arfsten, S.~Leupold, C.~Bradtm\"{o}ller, I.~Kampen, and A.~Kwade,
\newblock Colloids Surf. B: Biointerfaces {\bf 79}, 284 (2010).

\bibitem{Vella12}
D.~Vella, A.~Ajdari, A.~Vaziri, and A.~Boudaoud,
\newblock J. R. Soc. Interface {\bf 9}, 448 (2012).

\bibitem{gordon04}
V.~D. Gordon, X.Chen, J.~W. Hutchinson, A.~R. Bausch, M.~Marquez, and D.~A.
  Weitz,
\newblock J. Am. Chem. Soc. {\bf 126}, 14117 (2004).

\bibitem{Lee2008}
C.~Lee, X.~Wei, J.~W. Kysar, and J.~Hone,
\newblock Science {\bf 321}, 385 (2008).

\bibitem{CastellanosGomez2015}
A.~Castellanos-Gomez, V.~Singh, H.~S.~J. {van der Zant}, and G.~A. Steele,
\newblock Ann. Phys. {\bf 527}, 27 (2015).

\bibitem{Holmes10}
D.~P. Holmes and A.~J. Crosby,
\newblock Phys. Rev. Lett. {\bf 105}, 038303 (2010).

\bibitem{Vella11}
D.~Vella, A.~Ajdari, A.~Vaziri, and A.~Boudaoud,
\newblock Phys. Rev. Lett. {\bf 107}, 174301 (2011).

\bibitem{Vella15}
D.~Vella, J.~Huang, N.~Menon, T.~P. Russell, and B.~Davidovitch,
\newblock Phys. Rev. Lett. {\bf 114}, 014301 (2015).

\bibitem{Ripp18}
M.~M. Ripp, V.~D\'{e}mery, T.~Zhang, and J.~D. Paulsen,
\newblock arxiv:1804.02421 (2018).


\bibitem{Paulsen16}
J.~D. Paulsen, E.~Hohlfeld, H.~King, Z.~Qiu, T.~P. Russell, N.~Menon, D.~Vella,
  and B.~Davidovitch,
\newblock Proc. Natl Acad. Sci. USA {\bf 113}, 1144 (2016).

\bibitem{Box17}
F.~Box, D.~Vella, R.~W. Style, and J.~A. Neufeld,
\newblock Proc. R. Soc. A {\bf 473}, 20170335 (2017).

\bibitem{Cerda03}
E.~Cerda and L.~Mahadevan,
\newblock Phys. Rev. Lett. {\bf 90}, 074302 (2003).

\bibitem{Huang07}
J.~Huang, M.~Juszkiewicz, W.~H. \protect{de Jeu}, E.~Cerda, T.~Emrick,
  N.~Menon, and T.~P. Russell,
\newblock Science {\bf 317}, 650 (2007).

\bibitem{Davidovitch11}
B.~Davidovitch, R.~D. Schroll, D.~Vella, M.~Adda-Bedia, and E.~Cerda,
\newblock Proc. Natl. Acad. Sci. USA {\bf 108}, 18227 (2011).

\bibitem{Davidovitch12}
B.~Davidovitch, R.~D. Schroll, and E.~Cerda,
\newblock Phys. Rev. E {\bf 85}, 066115 (2012).

\bibitem{Taffetani17}
M.~Taffetani and D.~Vella,
\newblock Phil. Trans. R. Soc. A {\bf 375}, 20160330 (2017).

\bibitem{lo1983}
L.~L. Lo,
\newblock J. Fluid Mech. {\bf 132}, 65 (1983).

\bibitem{abramowitz64}
M.~Abramowitz and I.~A. Stegun,
\newblock {\em Handbook of Mathematical Functions with Formulas, Graphs, and
  Mathematical Tables},
\newblock Dover, New York, 1964.

\bibitem{Coman07}
C.~D. Coman and A.~P. Bassom,
\newblock J. Mech. Phys. Solids {\bf 55}, 1601 (2007).

\bibitem{Pipkin86}
A.~C. Pipkin,
\newblock IMA J. Appl. Math. {\bf 36}, 85 (1986).

\bibitem{Steigmann90}
D.~J. Steigmann,
\newblock Proc. R. Soc. A {\bf 429}, 141 (1990).

\bibitem{Vella17}
D.~Vella and B.~Davidovitch,
\newblock Soft Matter {\bf 13}, 2264 (2017).

\bibitem{Huang10}
J.~Huang, B.~Davidovitch, C.~D. Santangelo, T.~P. Russell, and N.~Menon,
\newblock Soft Matter {\bf 105}, 038302 (2010).

\bibitem{Vella15EPL}
D.~Vella, H.~Ebrahimi, A.~Vaziri, and B.~Davidovitch,
\newblock Europhys. Lett. {\bf 112}, 24007 (2015).

\bibitem{Chopin08}
J.~Chopin, D.~Vella, and A.~Boudaoud,
\newblock Proc. R. Soc. Lond. A {\bf 464}, 2887 (2008).

\bibitem{Hure12}
J.~Hure, B.~Roman, and J.~Bico,
\newblock Phys. Rev. Lett. {\bf 109}, 054302 (2012).

\bibitem{Paulsen17}
J.~D. Paulsen, V.~D\'{e}mery, K.~B. Toga, Z.~Qiu, T.~P. Russell,
  B.~Davidovitch, and N.~Menon,
\newblock Phys. Rev. Lett. {\bf 118}, 048004 (2017).

\bibitem{Taylor15}
M.~Taylor, B.~Davidovitch, Z.~Qiu, and K.~Bertoldi,
\newblock J. Mech. Phys. Sol. {\bf 79}, 92 (2015).

\bibitem{Bella14}
P.~Bella and R.~V. Kohn,
\newblock Commun. Pure Appl. Math. {\bf 67}, 693 (2014).

\bibitem{King12}
H.~King, R.~D. Schroll, B.~Davidovitch, and N.~Menon,
\newblock Proc. Natl. Acad. Sci. USA {\bf 109}, 9716 (2012).

\end{thebibliography}

\end{document}